\documentclass[aps,preprint,pdftex]{revtex4-1}
\usepackage{amssymb}
\usepackage{amsmath}
\usepackage{graphicx}
\usepackage{bm}
\usepackage{color}
\usepackage{mathrsfs}
\usepackage{subfigure}
\usepackage{caption}
\usepackage{ulem}
\usepackage{epstopdf}
\usepackage{amsmath}
\DeclareMathOperator{\sgn}{sgn}
\DeclareMathOperator{\Imag}{Im}
\begin{document}


\title{Analytical model of plasma response to external magnetic perturbation in absence of no-slip condition}
\author{Wenlong Huang}
\affiliation{School of Computer Science and Technology, Anhui University of Technology, Ma'anshan, Anhui 243002, China}

\author{Ping Zhu}
\email[E-mail:]{zhup@hust.edu.cn}
\affiliation{International Joint Research Laboratory of Magnetic Confinement Fusion and Plasma Physics, State Key Laboratory of Advanced Electromagnetic Engineering and Technology, School of Electrical and Electronic Engineering, Huazhong University of Science and Technology, Wuhan, Hubei 430074, China\\
Department of Engineering Physics, University of Wisconsin-Madison,
Madison, Wisconsin 53706, USA
}

\date{\today}

\begin{abstract}

Recent simulation and experimental results suggest that the magnetic island and flow on resonant surface often do not satisfy the ``no-slip" condition in the steady state. A new theory model on nonlinear plasma response to external magnetic perturbation in absence of no-slip condition is proposed. The model is composed of the equations for the evolution of both width and phase of magnetic island due to forced reconnection driven by the external magnetic perturbation, and the force-balance equation for the plasma flow. When the island width is much less than the resistive layer width, the island growth is governed by the linear Hahm-Kulsrud-Taylor solution in presence of time-dependent plasma flow. In the other regime when the island width is much larger than the resistive layer width, the evolution of both island width and phase can be described using the Rutherford theory. The island solution is used to construct the quasi-linear electromagnetic force, which together with viscous one, contributes to the nonlinear variation in plasma flow. The no-slip condition assumed in the conventional error field theory is not imposed here, where the island oscillation frequency depends on but does not necessarily equal to the plasma flow frequency at the rational surface.

\end{abstract}

\maketitle

\section{Introduction}
Resonant Magnetic Perturbations (RMPs) refer to the nonaxisymmetric magnetic field at tokamak boundary externally imposed by various coil systems. Traditionally, RMPs have been used as an effective tool to detect and compensate the error fields~\cite{rao13a, wang16a} in order to prevent model locking induced by these field errors~\cite{nave90}. RMPs have been also employed in tokamak devices to control MHD activities and mitigate major disruptions~\cite{liu10a, evans05a, sut11a, hu15a}. For example, in J-TEXT tokamak~\cite{ding2018, liang2019}, the core tearing mode can be suppressed or locked using the external RMPs by modulating RMP coil current amplitude and phase~\cite{hu12a, hu13a, jin15, hu16}.

The locking of tearing mode due to RMPs can be modeled using the error field (EF) theory~\cite{fitz93}. The EF model is composed of the modified Rutherford equation, the torque balance equation, and the no-slip condition. Based on the EF model, the dynamics of mode locking in presence of external coils and resistive wall is investigated~\cite{fitz01a}; and the scaling law of the error field penetration threshold as well as its extrapolation to ITER is obtained~\cite{fitz12a}. Recently, neoclassical and two-fluid effects are introduced to understand plasma response in high $\beta$ tokamak~\cite{fitz18a}. Using the EF model, we are able to explain many aspects of the mode locking and island suppression process in recent J-TEXT experiment as well as the locking-unlocking hysteresis phenomenon on EXTRAP-T$2$R~\cite{huang15a, huang16a}. However, the no-slip condition, assumed in the EF model may not be always valid. For example, EF model predicts that the plasma flow on resonant surface drops to zero in the steady state of plasma response to a static RMP~\cite{huang15a}, which does not agree with previous simulation where the plasma flow remains finite even as the plasma response approaches steady state~\cite{yu08b}. 

In 1985, Hahm and Kulsrud (HK)~\cite{hahm85} first investigated the forced magnetic reconnection in the Taylor problem. They showed that after the inertial regime, the forced reconnection evolves into the linear constant-$\psi$ resistive-inertial regime and then the nonlinear Rutherford regime. Later, both linear and nonlinear theory of forced magnetic reconnection are extended to rotating plasmas~\cite{fitz91a}. The HK solution for plasma response in presence of plasma flow is used to identify $11$ distinguishable response regimes, which are defined in terms of plasma viscosity, rotation, and resistivity~\cite{fitz98a}. These theories admit steady state solutions of forced reconnection in both linear, constant-$\psi$ and nonlinear Rutherford regimes, which may extend to the Sweet-Parker regime for larger boundary perturbation~\cite{wang92a, fitz03a, fitz04a, fitz04b}. Recently, plasmoid formation process driven by the boundary perturbation in the context of Taylor problem has been studied in theory~\cite{comisso15b, dewar13a, vek15a}.


The linear plasma response solution can be used to construct the quasi-linear forces. For example, using P\'ade approximation, the inner solution and the quasi-linear Maxwell and Reynolds forces are derived~\cite{cole15a}. Furthermore, qualitative agreement is achieved between NIMROD simulations and their theory results~\cite{beidler17a, beidler18a}. In all these linear and quasi-linear theories as well as the corresponding simulations, the no slip condition is neither required nor satisfied.


In this paper, we extend previous theories to model the plasma response to RMPs in absence of no-slip condition. The model is mainly composed of the magnetic response and the force balance equations. When the island width is much narrower than the resistive layer width, the island growth is governed by the extended HK solution in presence of plasma flow. Note that the plasma flow in our extended island solution can evolve with time, which is assumed constant in previous theory~\cite{fitz91a}. On the other hand, when the island is much wider than the resistive layer, the evolution of both island width and phase can be described using the Rutherford theory. The no-slip condition assumed in the conventional error field theory is not imposed here, where the island oscillation frequency depends on but does not necessarily equal to the plasma flow frequency at the rational surface. Our extended model is expected to agree better with recent simulations and experiments~\cite{yu08b, hu13a}

The rest of the paper is organized as follows. In Sec. II, we introduce the reduced MHD model of the Taylor problem. In Sec. III, we obtain the extended HK solution with time-dependent plasma flow to better understand recent simulations. Then we construct the quasi-linear forces and propose a new plasma response model in absence of no slip condition in Sec. IV. Finally, we give a summary and discussion in Sec. V.
\section{Model of the Taylor problem}

%
%

Introducing the flux function $\psi$ and the stream function $\phi$ in a Cartesian coordinate system, the magnetic field and the velocity can be written as $\vec{B}=B_T\vec{e}_z+\hat{z}\times\nabla\psi$ and $\vec{v}=\hat z\times\nabla\phi$. Then, the incompressible two field reduced MHD model governing $\psi$ and $F$ are given, respectively, by

\begin{eqnarray}
\frac{\partial\psi}{\partial t}+\vec{v}\cdot\nabla\psi=\frac{\eta}{\mu_0}\nabla^2\psi \label{rmhd1},\\
\rho(\frac{\partial}{\partial t}+\vec{v}\cdot\nabla)F=\vec{B}\cdot\nabla j_z + \nu_\perp \nabla^2 F \label{rmhd2},
\end{eqnarray}
where $F=\hat z\cdot\nabla\times\vec{v}=\nabla^2\phi$ is the vorticity and $j_z=\hat z\cdot\vec{j}=\frac{1}{\mu_0}\nabla^2\psi$, and $\rho$, $\eta$, and $\nu_\perp$ are the plasma density, resistivity, and viscosity, respectively. Hereafter, we only consider the small viscosity regime, in which the Prandtl number $P_r \ll 1$; and then, the viscous effect will be neglected unless otherwise stated. Based on the above Eqs. \eqref{rmhd1} and \eqref{rmhd2} without the viscous term, Hahm and Kulsrud (HK) first consider the Taylor problem in a static plasma~\cite{hahm85}.
In the Taylor problem, the plasma is surrounded by perfect conducting walls at $x=\pm a$. The equilibrium magnetic field is given as $\bm{B}=B_T\hat z+B_0\frac{x}{a}\hat y$, where $B_T$ and $B_0$ are all constants. Then the equilibrium flux function $\psi_{eq}=\frac{B_0}{2a}x^2$. The boundary perturbation is specified as $x=\pm(a-\delta_{\rm RMP} e^{iky})$, and the perturbed flux function assumes the form of $\psi_1=\psi_1(x)e^{iky}$. Here $\delta_{\rm RMP} = \delta_{\rm RMP} (t)$ is the amplitude of the boundary perturbation. Besides, the mirror symmetry $\psi_1(x)=\psi_1(-x)$ is also assumed.

\section{Linear solution of the Taylor problem in presence of plasma flow}

The linearized governing equations of $\psi_1$ and $\phi_1$ are
\begin{eqnarray}
&\frac{\partial}{\partial t} \psi_1 + \bm{v}_0 \cdot \nabla \psi_1 + \bm{v}_1 \cdot \nabla \psi_0 = \frac{\eta}{\mu_0} \nabla^2 \psi_1 \label{ln_psi},\\
&\rho (\frac{\partial}{\partial t} + \bm{v}_0 \cdot \nabla) F_1 + \rho \bm{v}_1 \cdot \nabla F_0 = \bm{B}_0 \cdot \nabla j_{z 1}\label{ln_phi},
\end{eqnarray}
where $\bm{v}_{0} = v_0(x, t) \hat y$, and $F_0 = \hat z \cdot \nabla \times \bm{v}_0 $. Note that $\bm{v}_0$ in Eqs. \eqref{ln_psi} and \eqref{ln_phi} can evolve with time, even when the island width is much less than the resistive layer width due to the strong modulation by magnetic perturbations~\cite{beidler17a, beidler18a}. To proceed, we divide the plasma rotation into two parts, i.e. $v_{0}(x, t) = v_{eq} + \delta v_0(x, t)$, where $v_{eq}$ is the constant equilibrium flow and $\delta v_0$ is the time-dependent part. We define $\psi_1 \equiv \hat\psi_1e^{-ik\delta\varphi_{\rm temp}(t)}$, $\phi_1 \equiv \hat\phi_1e^{-i\delta\varphi_{\rm temp}(t)}$, where $\delta\varphi_{\rm temp} = k\int_0^t\delta v_{0s}(t')dt'$ and $\delta v_{0s} = \delta v_0(0,t)$, $v_{0s} = v_{eq}+\delta v_{0s}$.


Neglecting the inertial and resistive terms, the outer solution to Eqs. \eqref{ln_psi} and \eqref{ln_phi} is
\begin{align}
\hat \psi_1 = \left\{\hat \psi_s [\cosh{(kx)} - \frac{\sinh{(kx)}}{\tanh{(ka)}}] + B_0  \frac{\sinh{(kx)}}{\sinh{(ka)}}\delta_{\rm RMP}(t) e^{i \delta\varphi_{\rm temp}(t)}\right\}e^{i k y} \label{hk_ln_o_ps},
\end{align}
where $\psi_s=\psi_1(0, t)$. In the inner region, by assuming $\partial / \partial_x \gg k$,  neglecting the flow shear, and introducing Laplace transform, we arrive at~\cite{hahm85}
\begin{eqnarray}
\frac{\partial^2}{\partial \bar \theta^2}\Psi=\bar \epsilon \bar \Omega(4\Psi+\bar \theta U)\label{lnps_in_lp},\\
\frac{\partial^2}{\partial \bar \theta^2}U-\frac14 \bar \theta^2 U=\bar \theta \Psi\label{lnph_in_lp},
\end{eqnarray}
where $\tilde\psi = \mathcal{L}[\hat\psi_1] = \displaystyle \int_0^\infty \hat\psi_1 e^{- s t} d t$, $\tilde \phi = \mathcal{L}[\hat\phi_1] = \displaystyle \int_0^\infty \hat\phi_1 e^{- s t} d t$, $\bar s=s + i k v_{eq}$,  $\bar \epsilon^4 = \frac{\bar s \tau_A^2}{4 (k a)^2 \tau_R}$, $\bar \nu = \frac{-i\bar s}{4 \bar \epsilon k^2}$, $\bar \Omega = \frac{\bar \epsilon \tau_R \bar s}{4}$, $\bar \theta = x / \bar \epsilon a$, $\Psi = k / B_0 \tilde \psi_1$, and $U = - \tilde \phi / \bar \nu$. Eliminating $U$ from Eq. \eqref{lnps_in_lp} by means of Eq. \eqref{lnph_in_lp} and introducing $Z=\frac{\partial^2}{\partial \bar \theta^2}\Psi$, we have
\begin{align}
\frac{d^3Z}{d \theta_1^3} = (\bar \mu + \theta_1^2) \frac{d Z}{d \theta_1} +4 \theta_1 Z \label{ln_in_Z},
\end{align}
where $\theta_1 = \bar \theta/ \sqrt2$ and $\bar \mu = 8 \bar \epsilon \bar \Omega$.

In this section, we study the plasma response to two types of transient or dynamic RMPs. The first type of the boundary perturbation is $\delta_{\rm RMP} = \delta_0 e^{-i\Omega t}$, where $\Omega$ is the rotating frequency of boundary perturbation. The second type of transient boundary perturbation takes the form
\begin{align}
&\delta_{\rm RMP}=\delta_0T_{\rm NIM}(t,\tau_0)+\delta_1T_{\rm NIM}(t',\tau_T)H(t')-\delta_1T_{\rm NIM}(t'',\tau_T)H(t''), \label{transRMPb}\\
&T_{\rm NIM}(t,\tau_0)=1-e^{-t/\tau_0}-\frac{t}{\tau_0}e^{-t/\tau_0}\label{transRMPa}.
\end{align}
Here $t'=t-t_T$, $t''=t'-\Delta t_T$, and $\delta_0$, $\delta_1$, $\tau_0$, $\tau_T$ and $t_T$ are all constants. Both types of RMPs have been considered in previous studies~\cite{fitz91a, beidler18a}, and they are adopted below in subsections $A-B$ and $C$ respectively for the linear solutions of the Taylor problem in presence of plasma flow.
\subsection{Plasma response to the first type of transient RMP in inertial regime}
Before we discuss the linear solution with constant-$\psi$ assumption, it is useful to study the island evolution in the inertial regime, where the constant-$\psi$ assumption is not valid. In the inertial regime, i.e. $t \ll \tau_R^{1/3} \tau_A^{2/3}$, the perturbed current term $\frac{d^3Z}{d \theta_1^3}$ in Eq. \eqref{ln_in_Z} can be neglected. We note that the plasma rotation cannot be modulated by RMPs in such a time scale, and therefore the effect of finite $\delta v_0$ is neglected here. Combined with the boundary condition $Z(0) = \bar \mu/2 \Psi(0)$, $Z'(0) = 0$, and $Z(\infty) = 0$~\cite{hahm85}, the inner solution is obtained as
\begin{align}
Z=\frac{\bar \mu \Psi(0)}{2(1+\theta_1^2/\bar \mu)^2}.
\end{align}
In presence of plasma flow, the perturbed flux at the rational surface can be approximated through asymptotic matching as
\begin{align}
\Psi = \frac{4}{\pi} \frac{k^3 a^2 \delta_0}{\sinh{(k a)}} \frac{1}{\tau_A \tau_R (s+i\Omega) \bar s^2}\label{ln_ps_ncstps_lps}.
\end{align}
Using the inverse Laplace transform, we obtain
\begin{align}
&\psi_s (t) = \frac{4}{\pi} \frac{B_0 \delta_0 k^2 a^2}{\sinh{(k a)} \tau_R \tau_A}\frac{e^{- i \omega t} + i \omega t e^{- i \omega t} - 1}{\omega^2} e^{- i \Omega t}\nonumber\\
&\qquad = \frac{4}{\pi} \frac{B_0 \delta_0 k^2 a^2}{\sinh{(k a)}}(\frac{\tau_A}{\tau_R })^{\frac13}\frac{e^{- i \hat\omega \hat t} + i \hat\omega \hat t e^{- i \hat\omega \hat t} - 1}{\hat\omega^2} e^{- i \hat\Omega \hat t}  \label{ln_ps_ncstps_lps},
\end{align}
where $\omega = k v_{eq} -\Omega$, $\hat\omega=\omega\tau_A^{2/3}\tau_R^{1/3}$, $\hat\Omega=\Omega\tau_A^{2/3}\tau_R^{1/3}$, and $\hat t= \frac{t}{\tau_A^{2/3}\tau_R^{1/3}}$. It is worth noting that the perturbed flux at the rational surface cannot be written as $\psi_s = |\psi_s|e^{-i \varphi_s}$, where $\varphi_s= \omega t + \varphi_{s0}$ is the phase of $\psi_s$. Thus the no-slip condition cannot be satisfied here, which would require that the perturbed flux satisfy $\psi_s(t) = |\psi_s(t)| e^{-i \omega t}$ in the linear regime~\cite{fitz93a}. In addition, Eq. \eqref{ln_ps_ncstps_lps} can reduce to the linear solution in HK theory for static plasma when $\Omega = 0$.

Due to the time scale we consider here is $t \ll \tau_R^{1/3} \tau_A^{2/3}$, or, $\hat t \ll 1$, we can expand $e^{- i \hat\omega \hat t}$ as $e^{- i \hat \omega \hat t} \approx 1 - i \hat \omega \hat t + \frac12 (-i \hat \omega \hat t)^2 + \frac{1}{6} (-i\hat \omega \hat t)^3$ in typical plasma regimes, and Eq. \eqref{ln_ps_ncstps_lps} can be approximated as
\begin{align}
\psi_s (t) \approx \frac{2}{\pi} \frac{B_0 \delta_0 k^2 a^2}{\sinh{(k a)}}(\frac{\tau_A}{\tau_R})^{1/3} (\hat t^2 - \frac23 i \hat \omega \hat t^3)e^{- i \hat \Omega \hat t} \approx \frac{2}{\pi} \frac{B_0 \delta_0 k^2 a^2}{\sinh{(k a)}} (\frac{\tau_A}{\tau_R})^{1/3} (\hat t^2 - \frac23 i \hat \omega \hat t^3 - i \hat \Omega \hat t^3).
\end{align}
From the above expression of the perturbed flux, we find that the real part of $\psi_s$ grows in $t^2$, which is exactly same as the HK theory. Both plasma and RMP rotations (i.e. $v_{eq}$ and $\Omega$) can introduce an imaginary part to the perturbed flux, which grows in $t^3$.
%
\subsection{Plasma response to the first type of transient RMP in resistive-inertial regime}
When $t \gg \tau_R^{\frac13} \tau_A^{\frac23}$, the island evolves into the resistive-inertial regime, where the constant-$\psi$ approximation is valid for an appropriate amplitude of the boundary perturbation~\cite{wang92a, comisso15b}. Plasma flow can be strongly modulated by the boundary perturbations in such a time scale~\cite{beidler17a, beidler18a}. Thus the time-dependent part of plasma flow $\delta v_0$ should be included in the resistive-inertial regime. Expanding $U = \Sigma a_n \hat D_n$ as in ~\cite{furth63a}, we obtain the expression for the expanding coefficients from Eq. \eqref{lnph_in_lp}
\begin{align}
a_n = - \frac{\int \bar \theta \hat D_n d\bar \theta}{n+\frac12}\Psi\label{phi_an},
\end{align}
where
\begin{align}
\hat D_n(\theta) = (-1)^n \frac{e^{\frac{\theta^2}{4}}}{(2 \pi)^{\frac14}(n!)^{\frac12}}\frac{d^n}{d \theta^n}e^{\frac{-\theta^2}{2}},
\end{align}
is the normalized parabolic cylinder function. After the asymptotic matching, we arrive at
\begin{align}
\frac{12 \bar \Omega}{a} \Psi = \Delta_0'\Psi + \Delta_e'\Psi_c\label{HKLaplace},
\end{align}
where $\Delta_0'$ and $\Delta_e'$ are the jump in $d\ln{\psi_1}/dx$ across the rational surface and the tearing drive due to the external field, and $\Psi_c = \mathcal{L}[k\delta_{\rm RMP}(t)e^{i\delta\varphi_{\rm temp}(t)}]$. Dividing $\frac{\bar\Omega}{s+ikv_{eq}}$ in both sides of Eq. \eqref{HKLaplace}, and applying the inverse Laplace transform, we rearrange the above equation as
\begin{align}
&\frac{\tau_\delta}{a}[\frac{\partial}{\partial t}+ikv_{0s}(t)]\psi_s =& \nonumber\\
&\qquad  \left\{\Delta_0'\int_0^tG_1(t-t')\psi_s(t')e^{i\varphi_{\rm temp}(t')}dt' + \Delta_e'\int_0^tG_1(t-t')\psi_c(t')e^{i\varphi_{\rm temp}(t')}dt'\right\}e^{-i\varphi_{\rm temp}(t)}\label{EQ_HKwflow},\\
&G_1(t) = \frac{\Gamma(\frac34)}{\sqrt{2}\pi}(ikv_{eq})^{\frac14}t^{-\frac34},\\
&\tau_\delta = \frac{3}{\sqrt{2}}\tau_R^{\frac34}\tau_A^{\frac12}|kv_{eq}|^{\frac14}e^{i\frac{\pi}{8}\sgn{(kv_{eq})}}/(ka)^{\frac12},
\end{align}
where $\psi_c (t) = B_0\delta_{\rm RMP}(t)$, $\varphi_{\rm temp}(t) = k\int_0^tv_{0s}(t')dt'$, $\tau_\delta$ is the layer response time in the resistive-inertial regime~\cite{fitz98a}, and $\sgn{(kv_{eq})}$ is the sign of $kv_{eq}$.

On the other hand, Eq. \eqref{HKLaplace} can be also rearranged as
\begin{align}
\Psi(0)=\frac{\Psi_c}{\cosh{(k a)} + [3 \sinh{(k a)} /(2 k a)^{\frac32}] \bar p^{\frac54}},
\end{align}
where $\bar p = \bar s \tau_R^{\frac35} \tau_A^{\frac25}$. Using the inverse Laplace transform, we obtain a transparent expression of the perturbed flux at the rational surface
\begin{align}
&\psi_s(t)= \frac{B_0}{\cosh{(ka)}}e^{-i\varphi_{\rm temp}(t)}\int_0^tG_2(t-t')\delta_{\rm RMP}(t')e^{i\varphi_{\rm temp}(t')}dt'\label{newHK},\\
&G_2(t) = \tau_R^{\frac35}\tau_A^{\frac25}\times\left\{-\frac45[P_Ae^{P_A \tau}  + P_Be^{P_B \tau}] - \frac{\lambda }{\sqrt{2}\pi} \displaystyle \int_0^\infty e^{- u \tau} \frac{u^{\frac54}}{(1 - \sqrt{2} \lambda u^{\frac54} + \lambda^2 u^{\frac52})} d u\right\},
\end{align}
where $\lambda = 3/2^{\frac32} \tanh{(k a)}/(k a)^{\frac32}$, $\tau = t/\tau_R^{\frac35} \tau_A^{\frac25}$, and $P_{A,B}=\lambda^{-\frac45}exp(\pm\frac{4 \pi i}{5})$. Eq. \eqref{newHK} should be the solution of Eq. \eqref{EQ_HKwflow}. It is worth noting that the plasma flow in Eq. \eqref{newHK} can depend on time, which is different from the constant flow assumed in previous extended HK theory~\cite{fitz91a}. We also note that Eq. \eqref{newHK} cannot be directly written as $\psi_s = |\psi_s|e^{-i\varphi_{\rm temp}}$, which means that the no-slip condition is also invalid in this regime.

To understand recent simulation results~\cite{becoulet12a, furukawa09a}, we neglect the time-dependent part of plasma flow $\delta v_0$ and focus on the plasma response to the first type of transient RMP boundary perturbation, i.e. $\delta_{\rm RMP} = \delta_0 e^{- i \Omega t}$. The expression of $\psi_s$ in Eq. \eqref{newHK} reduces to
\begin{align}
&\psi_s(t) =  \frac{B_0 \delta_{\rm RMP}}{\cosh{(ka)}}[\frac{1}{1+\lambda(i R)^{\frac54}} \nonumber\\
&\qquad - \frac45 e^{- i \omega t}(\frac{P_A}{P_A - i R}e^{P_A \tau}  + \frac{P_B}{P_B - i R}e^{P_B \tau}) \nonumber\\
&\qquad + \frac{\lambda }{\sqrt{2}\pi} e^{- i \omega t} \displaystyle \int_0^\infty e^{- u \tau} \frac{u^{\frac54}}{(u + i R)(1 - \sqrt{2} \lambda u^{\frac54} + \lambda^2 u^{\frac52})} d u],\label{ps_evo_RI}
\end{align}
where $R = \omega \tau_R^{\frac35} \tau_A^{\frac25}$. The linear solution of plasma response in Eq. \eqref{ps_evo_RI} is consistent with previous result in cylindrical configuration~\cite{fitz91a}. Due to the time oscillation from RMP, the first term in Eq. \eqref{ps_evo_RI} purely oscillates in time, whereas other terms may oscillate and grow in time. Eventually, the perturbed flux evolves to a steady state in the frame of $v = \Omega / k$.

\begin{figure}[htbp]
\setlength{\unitlength}{0.5cm}
\begin{center}
\includegraphics[width= 10 cm]{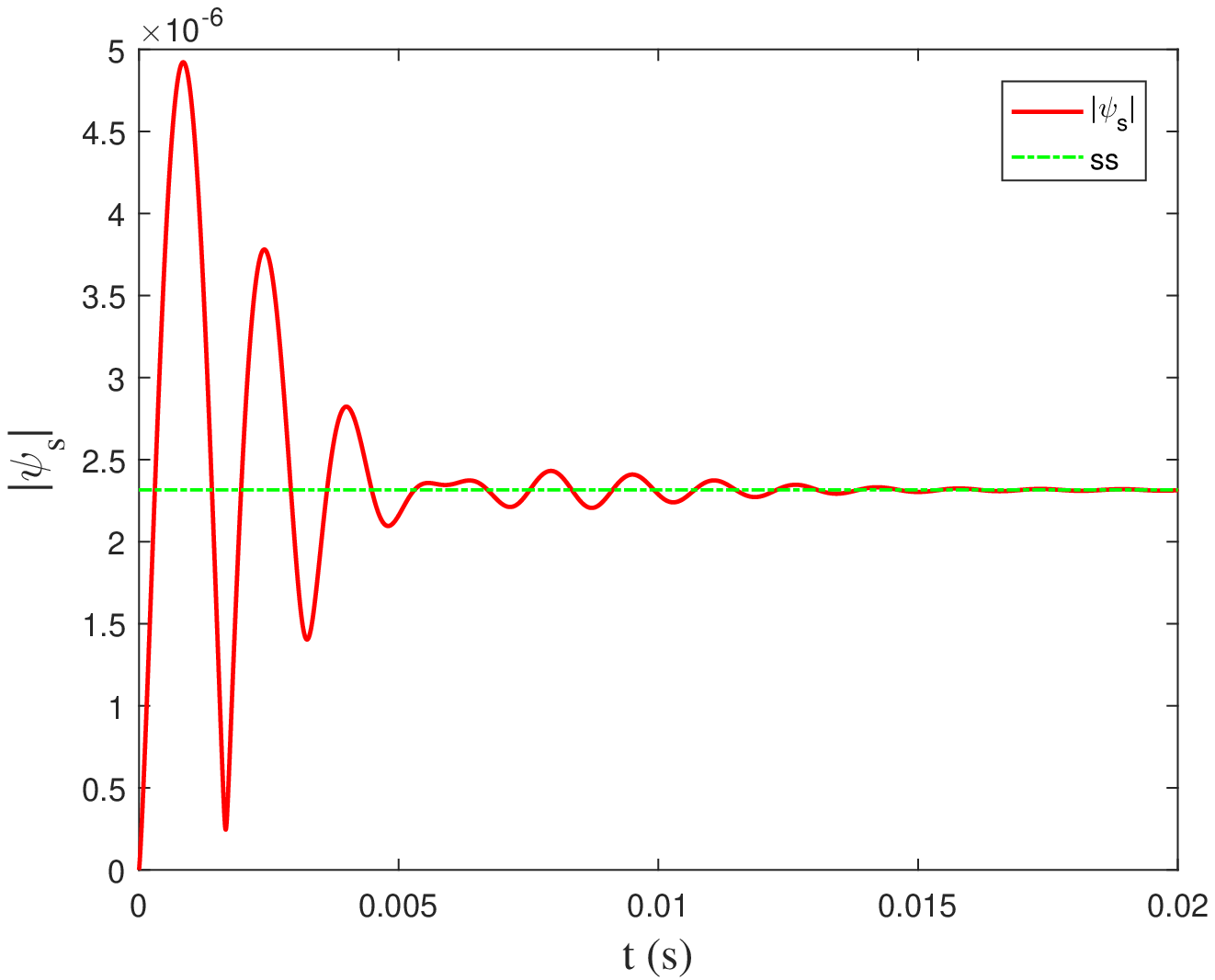}\\
\includegraphics[width= 10 cm]{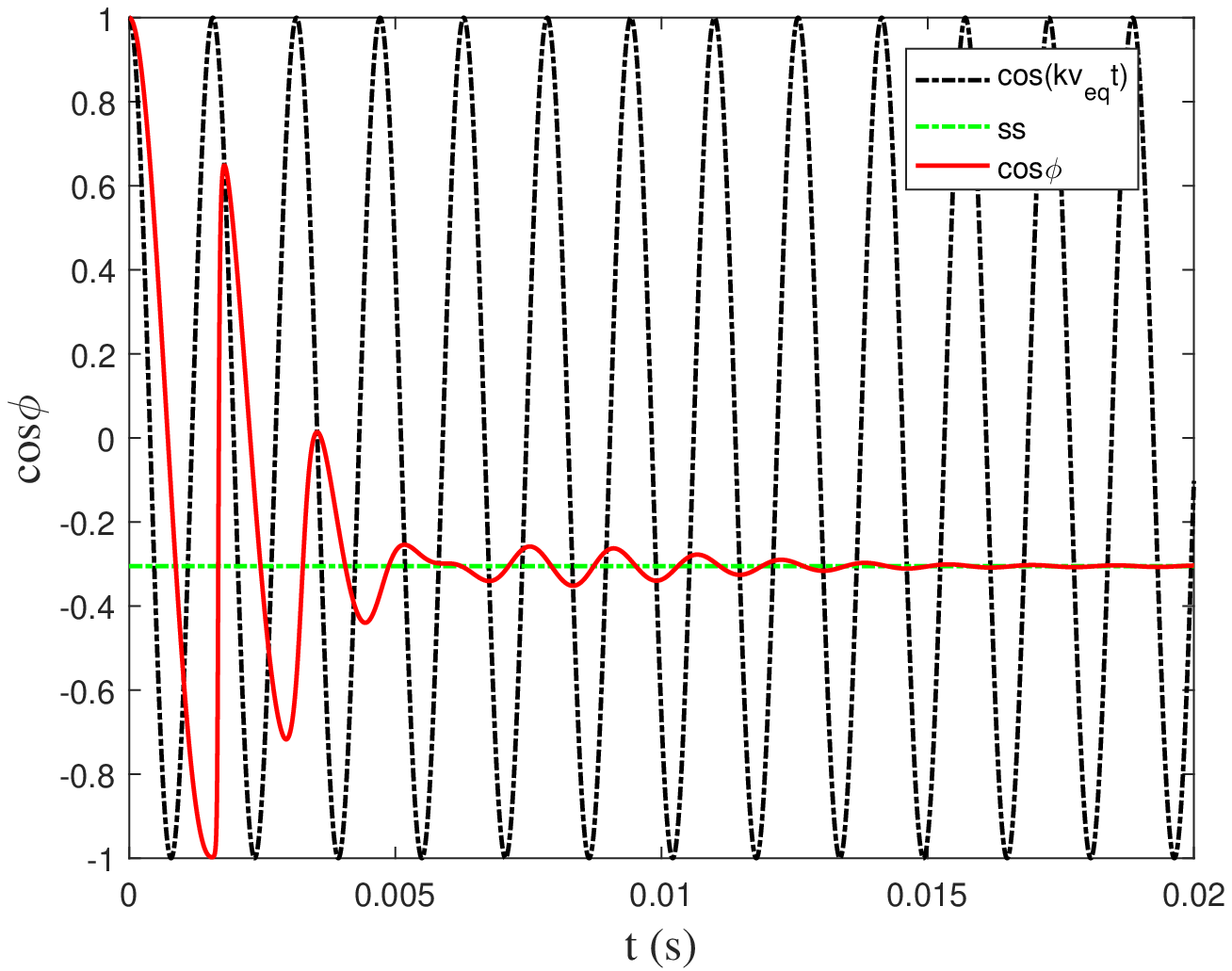}
\caption{The island width (upper) and phase (lower) as functions of time from Eq. \eqref{ps_evo_RI}. Parameters used here are $\eta=10^{-7}\Omega m$ and $v_{eq}=2000m/s$. The black dashed curve in the lower panel represents $\cos{(kv_{eq}t)}$. The green horizontal line in both panels represent the corresponding island width and phase in steady state.}
\label{fig2}
\end{center}
\end{figure}

The linear solution in Eq.\eqref{ps_evo_RI}, though obtained only in the resistive inertial regime, may be able to account for many features of plasma response found in simulation results~\cite{beidler17a, furukawa09a, becoulet12a}. To illustrate this, we examine the time evolution of the amplitude and phase of the magnetic island driven of the static boundary magnetic perturbation as predicted in Eq. \eqref{ps_evo_RI} (Fig. \ref{fig2}). The basic parameters used here are $a=0.5m$, $k=1/a$, $\rho=1.67\times 10^{-8}Kg/m^3$, $B_0=0.2T$, $B_T=2T$, and $\delta_0=2\times 10^{-4}m$. The island width oscillates and increases to a final steady state, which agrees with the above discussion. Furthermore, $\cos {\varphi}$ dramatically deviates from $\cos{(k v_{eq} t)}$, which does not satisfy the no-slip condition~\cite{fitz93a}.

\begin{figure}[htbp]
\setlength{\unitlength}{0.5cm}
\begin{center}
\includegraphics[width= 10 cm]{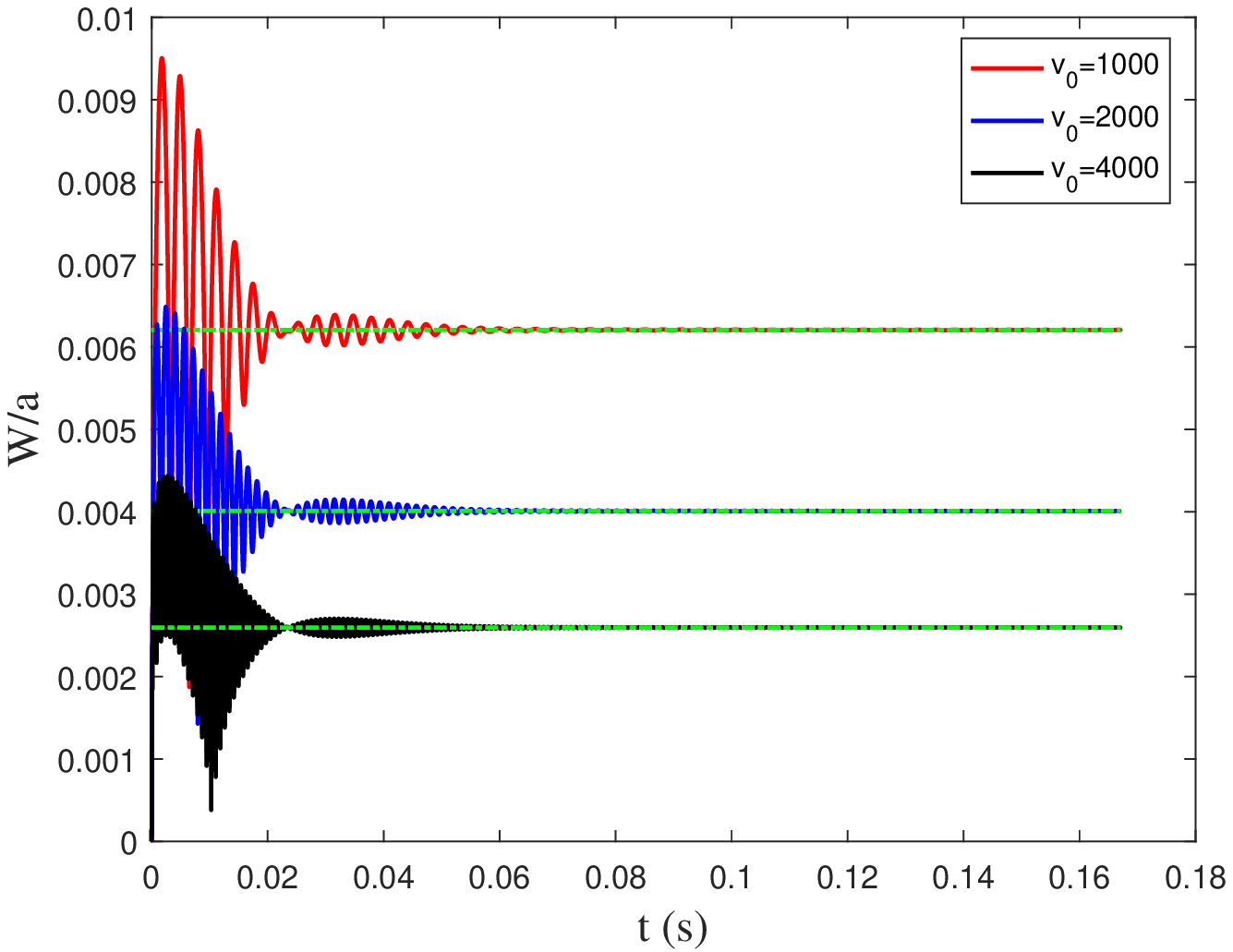}\\
\includegraphics[width= 10 cm]{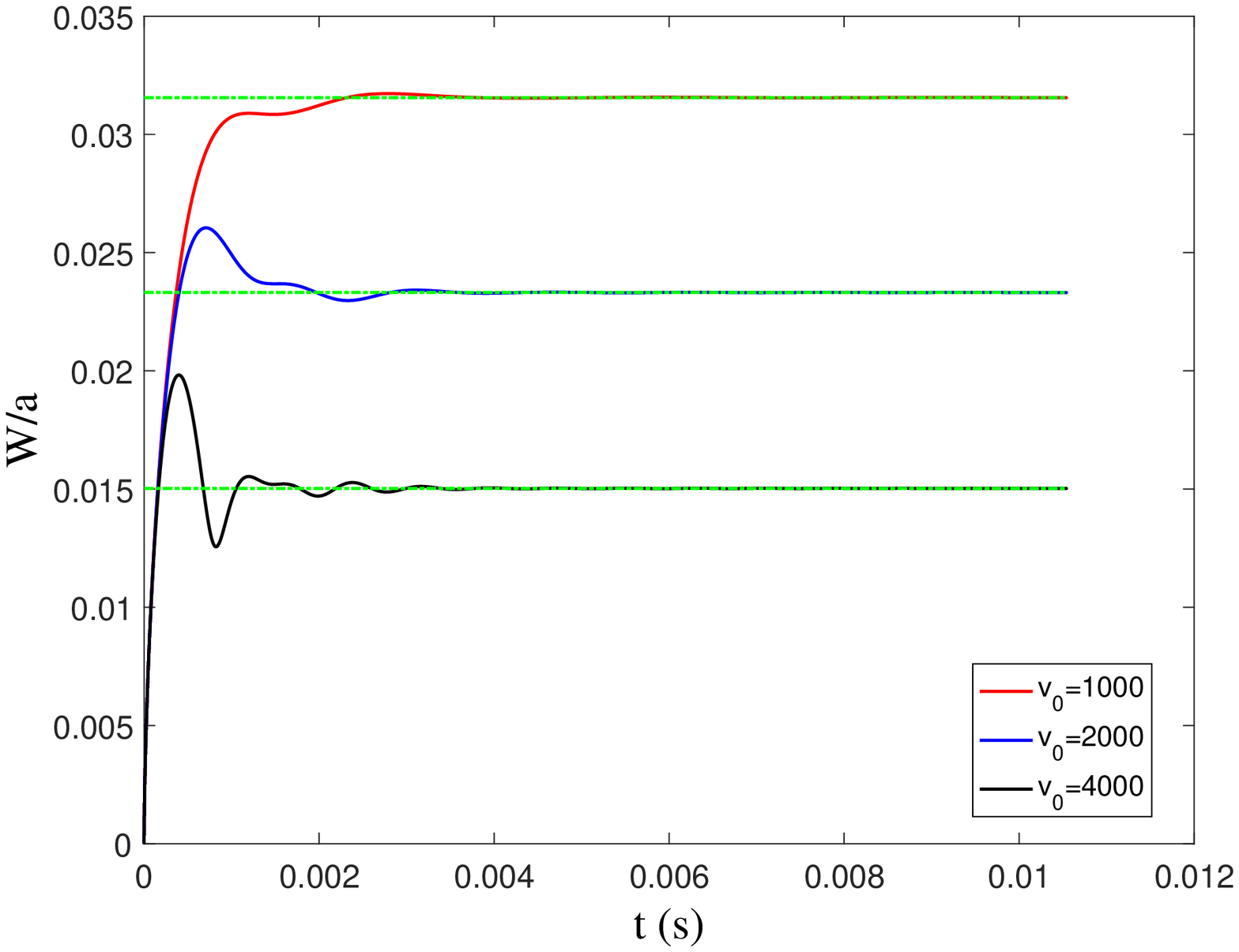}
\caption{The island width as a function of time from Eq. \eqref{ps_evo_RI} in different parameter regimes.  The plasma resistivities used here are $\eta=10^{-8}\Omega m$ (upper) and $\eta=10^{-6}\Omega m$ (lower), respectively. The red, blue and black curves represent $v_{eq}=1000m/s$, $v_{eq}=2000m/s$, and $v_{eq}=4000m/s$, respectively. The green horizontal lines represent the island width in steady state.
}
\label{fig3}
\end{center}
\end{figure}

The linear solution in Eq. \eqref{ps_evo_RI} also predicts the flow screening effects on plasma response in different resistive regimes. For example, in the upper panel of Fig. \ref{fig3}, where the resistivity is relative small ($\eta = 1 \times 10^{- 8}\Omega m$), we find that the island width oscillates and increases in time before reaching a steady state, and the oscillation frequency increases with the flow speed. In the lower panel, where the resistivity is relatively larger ($\eta = 1 \times 10^{- 6}\Omega m$), the oscillation in island growth almost disappears for the same plasma flow as in the upper panel. Both panels in Fig. \ref{fig3} show that the island width in steady state decreases with the plasma flow, demonstrating the flow shielding effect (See also Fig. \ref{fig5}).



Previous simulations find that the island width increases as well as oscillates before the final penetration state, similar to those shown in the upper panel of Fig. \ref{fig3}~\cite{yu08b, hu12a}. On the other hand, recent NIMROD simulations find the RMP induced island growth resembles those shown in the lower panel of Fig. \ref{fig3}, where the oscillation in island growth is weak or absent~\cite{beidler17a, beidler18a}. Our results in Fig. \ref{fig3} suggest that such a difference in island growth may attribute to different plasma parameters such as the resistivity involved.

\begin{figure}[htbp]
\setlength{\unitlength}{0.5cm}
\begin{center}
\includegraphics[width= 10 cm]{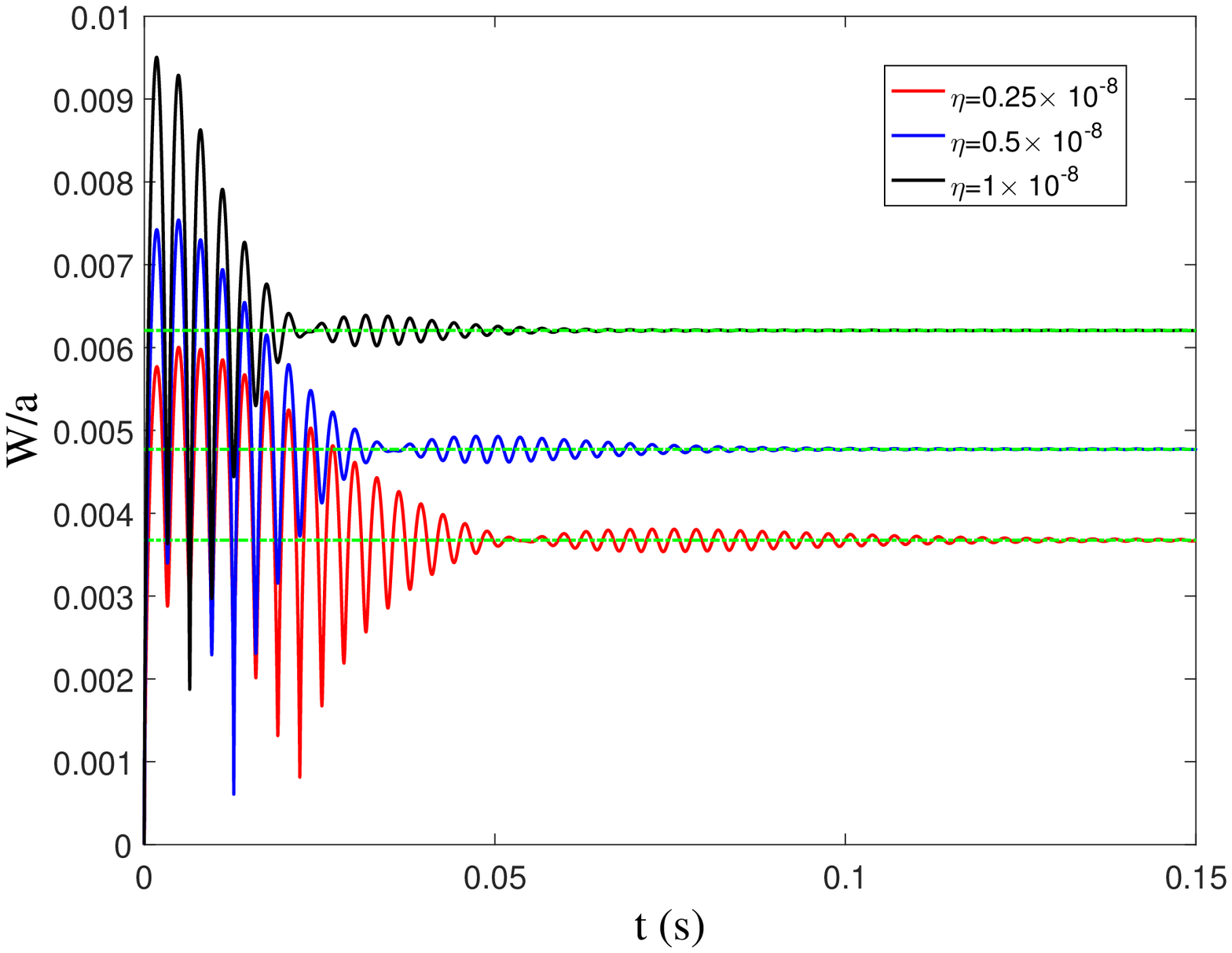}\\
\includegraphics[width= 10 cm]{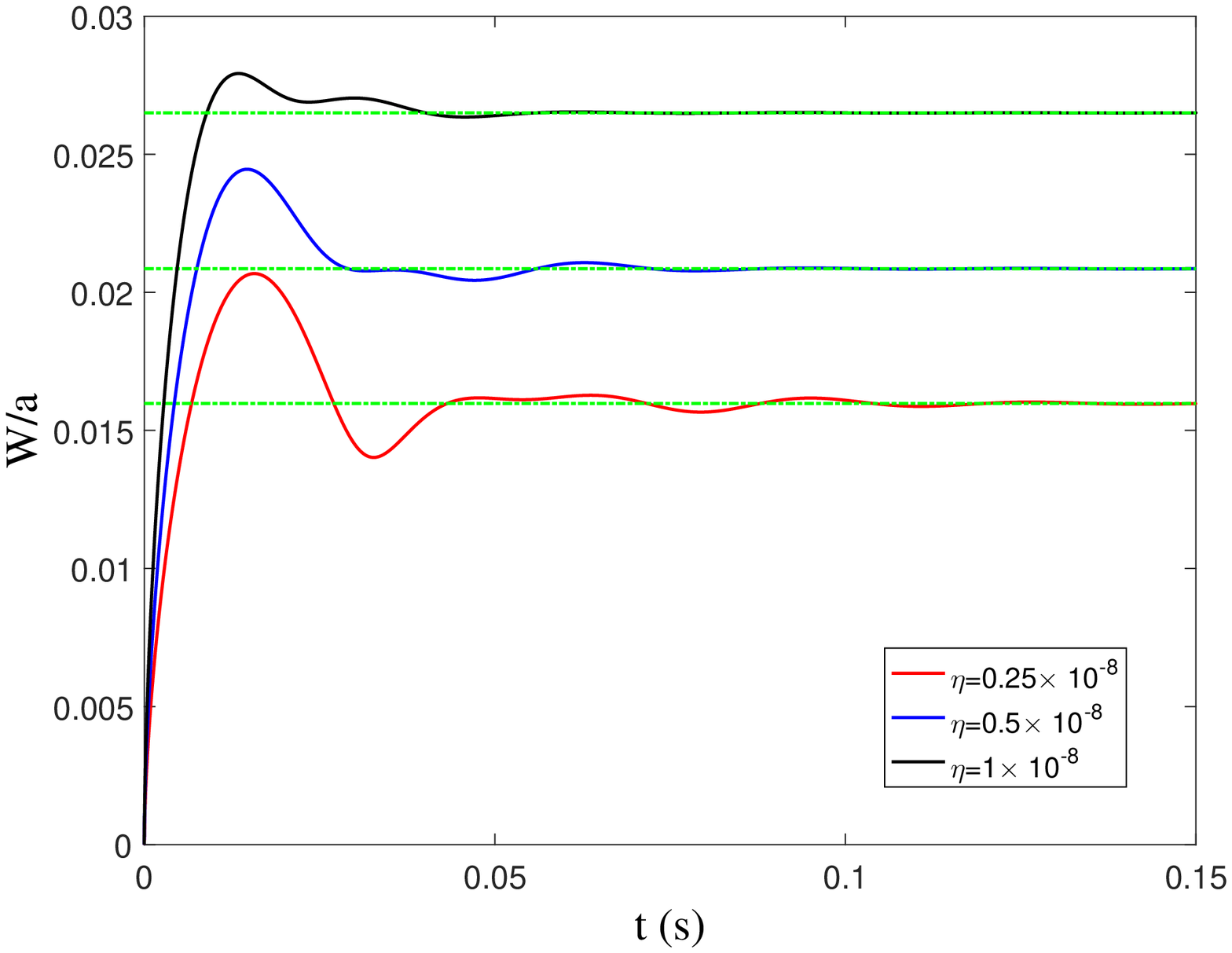}
\caption{The island width as a function of time from Eq. \eqref{ps_evo_RI} in different parameter regimes. The plasma flows used here are $v_{eq}=1000m/s$ (upper) and $v_{eq}=100m/s$ (lower), respectively. The red, blue, and black curves represent $\eta=0.25\times 10^{-8}\Omega m$, $\eta=0.5\times 10^{-8}\Omega m$, and $\eta=1\times 10^{-8}\Omega m$, respectively.}
\label{fig4}
\end{center}
\end{figure}

\begin{figure}[htbp]
\setlength{\unitlength}{0.5cm}
\begin{center}
\includegraphics[width= 10 cm]{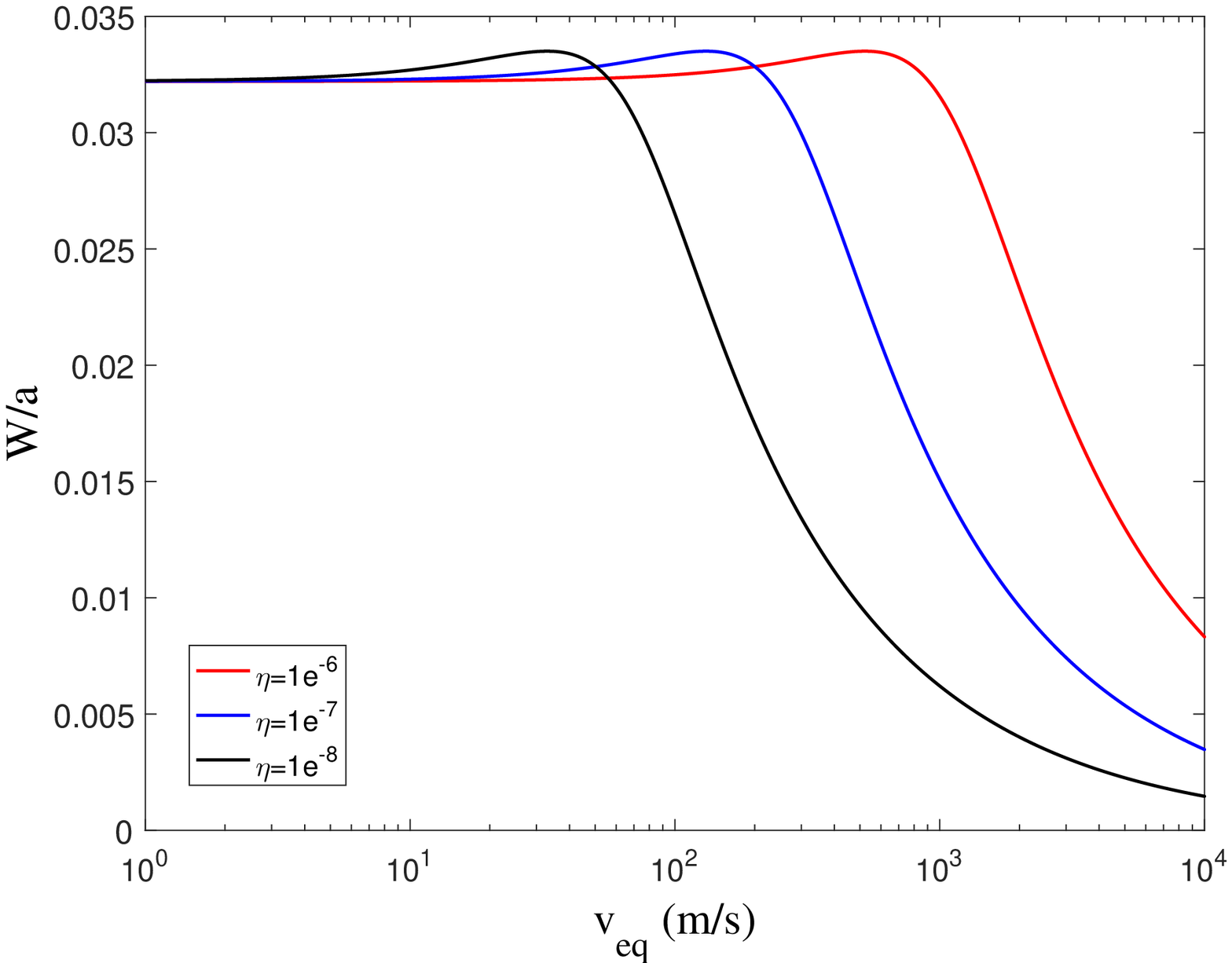}\\
\includegraphics[width= 10 cm]{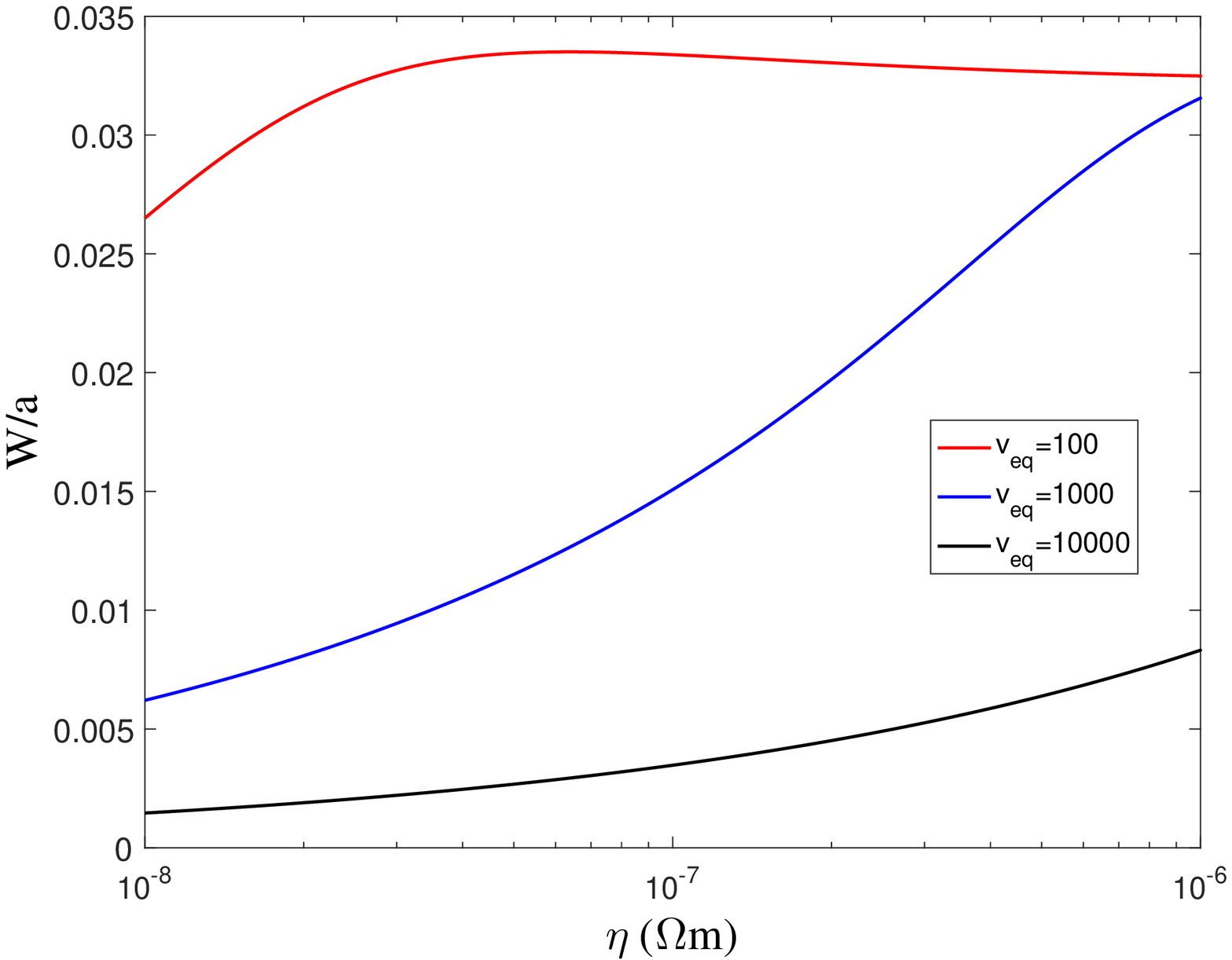}
\caption{Dependence of island width in steady state on the plasma flow (upper) and resistivity (lower) for different $\eta$ (upper) and $v_{eq}$ (lower). The red, blue, and black curves in the upper (lower) panel represent $\eta=10^{-6}\Omega m$ ($v_{eq}=100m/s$), $\eta=10^{-7}\Omega m$ ($v_{eq}=1000m/s$), and $\eta=10^{-8}\Omega m$ ($v_{eq}=10000m/s$), respectively.}
\label{fig5}
\end{center}
\end{figure}

The effects of plasma resistivity on RMP penetration process can be further illustrated from the linear solution in Eq. \eqref{ps_evo_RI} plotted in Fig. \ref{fig4}. In the upper panel of Fig. \ref{fig4}, one find that the plasma response oscillates and increases to steady state for a larger plasma flow ($v_{eq}=1000m/s$).  For a lower plasma flow speed ($v_{eq}=100m/s$), the oscillation in island growth nearly vanishes. Both panels of Fig. \ref{fig4} show that the island width in steady state increases with plasma resistivity, which qualitatively agrees with Ref.~\cite{becoulet12a} and also will be displayed in the lower panel of Fig. \ref{fig5}.

 To better understand the features of steady state shown in Figs. \ref{fig3} and \ref{fig4}, we derive from Eq. \eqref{ps_evo_RI} the analytical expressions for the steady island width and phase at $t \rightarrow \infty$ as
\begin{align}
& W = \frac{W_0}{\sqrt{1 + \lambda^2 |R|^{\frac52} + 2 \lambda |R|^{\frac54} \cos {\frac{5 \pi}{8}}}} \label{anltW}, \\
& \sin{ \varphi} = - \frac{\lambda |R|^{\frac54} \sin {\frac{5 \pi}{8}} \sgn{(kv_{eq})}}{1 + \lambda^2 |R|^{\frac52} + 2 \lambda |R|^{\frac54} \cos {\frac{5 \pi}{8}}}\label{anltP},
\end{align}
where $\varphi$ is the island phase. Here $W=2\sqrt{\frac{a}{B_0}|\psi_s|}$ is the island width and $W_0 = 2 \sqrt{\frac{a \delta_{0}}{\cosh{(k a)}}}$. From Eq. \eqref{anltW}, the island width increases with $|R|$ if $|R| \ll 1$ due to the fact that $\cos{\frac{5 \pi}{8}} < 0$. Therefore, the steady island width increases with plasma flow in the small $v_{eq}$ regime when $\eta$ is fixed, whereas it decreases with plasma resistivity in the large $\eta$ regime with a fixed $v_{eq}$. On the other hand, $W \propto \frac{1}{|R|^{5/4}}$ when $|R| \gg 1$. Accordingly, the steady island width decreases with plasma flow in the large $\eta$ regime with a fixed $v_{eq}$, and it increases with plasma resistivity in small $v_{eq}$ regime when $\eta$ is fixed. The above description is consistent with Fig. \ref{fig5}. In addition, the island phase $\phi \sim 0$ if $|R| \ll 1$, and $\phi \sim \pm \frac{5 \pi}{8}$ when $|R| \gg 1$. The $\pm$ is determined by the sign of $kv_{eq}$.

The dependence of the steady island width on the plasma flow (for different $\eta$) and the plasma resistivity (for different $v_{eq}$) governed by Eq. \eqref{anltW} and \eqref{anltP} can be further illustrated in Fig. \ref{fig5}. The upper panel of Fig. \ref{fig5} indicates that there is a threshold in plasma flow for any given resistivity. Below the threshold, the island width increases with the plasma flow in small $v_{eq}$ regime ($R \ll 1$). Above the threshold, the island width can be strongly shielded by plasma flow ($R \gg 1$), which is qualitatively consistent with the linear simulation result in Ref. \onlinecite{furukawa09a}. Such a threshold increases with plasma resistivity. In the lower panel of Fig. \ref{fig5}, the steady width of the forced island generally increases with resistivity except in the slower flow case ($v_{eq} = 100 m/s$), where the island width decreases with resistivity when $\eta$ becomes sufficiently large. Similar result can be also found in Fig. B1 of Ref. \onlinecite{furukawa09a}.

\subsection{Plasma response to the second type of transient RMP in resistive-inertial regime}
Plasma response to the second type of transient boundary perturbation as specified in Eqs. \eqref{transRMPa} and \eqref{transRMPb} has been previously studied using the NIMROD simulation along with an error field model in the following equation,
\begin{align}
[\frac{d}{dt'} + i\omega_{\rm res}(t')]B_{\rm res}(t') = \frac{a\Delta'_0}{\tau_\delta}B_{\rm res}(t') + \frac{a\Delta'_e}{\tau_\delta}B_{\rm ext}(t')\label{beidl},
\end{align}
where $\omega_{res} = \bm{k}\cdot\bm{v}_0(x=0)$, $B_{\rm res} = B_{x,1}(x=0)$, $B_{\rm ext} = B_{x,1}(|x|=a)$, and $\tau_\delta$ is the layer response time. The rest of definitions are conventional and can be found following Eq. ($17$) of Ref. \cite{beidler18a}. However, as we show here, such a model equation in Eq. \eqref{beidl} is only an approximation to the more rigorous flux evolution equation, i.e. Eq. \eqref{EQ_HKwflow}, in presence of time-dependent flow. In particular, when $s \ll ikv_{eq}$, $3\bar \epsilon \tau_R \approx \tau_\delta$. Using the inverse Laplace transform, both sides of Eq. \eqref{HKLaplace} can be rewritten, i.e. $(s + ikv_{eq})\frac{3\bar \epsilon \tau_R}{a}\Psi \rightarrow \frac{\tau_\delta}{a}[(\frac{\partial}{\partial t} + ikv_0)\psi_s]e^{i\delta\varphi_{\rm temp}(t)}$ and $\Delta_0'\Psi + \Delta_e'\Psi_c \rightarrow \Delta_0'\psi_se^{i\delta\varphi_{\rm temp}(t)} + \Delta_e'\psi_ce^{i\delta\varphi_{\rm temp}(t)}$. Then, Eq. \eqref{beidl} is thus obtained, which maybe appropriate only in the quasi-steady state when $s \ll ikv_{eq}$.

For second type boundary perturbation, if we neglect $\delta v_0(t)$ and keep only the constant equilibrium flow, the linear solution of plasma response in the resistive-inertial regime can be derived as
\begin{align}
\psi_1(0,t)=\frac{B_0\delta_0}{\cosh{(ka)}}\sum_{l=1,2,3}H_l(t,\tau_0)+\frac{B_0\delta_1}{\cosh{(ka)}}\sum_{l=1,2,3}[H_l(t',\tau_T)-H_l(t'',\tau_T)]\label{beidl18},
\end{align}
where
\begin{align}
&\small{H_1(t,\tau_j) = \frac{1}{1+\lambda(iR)^{\frac54}}-\frac{1}{1+\lambda(iR-q_j)^{\frac54}}e^{-t/\tau_j}-\frac{1}{1+\lambda(iR-q_j)^{\frac54}}\frac{2t}{\tau_j}e^{-t/\tau_j},}\\
&\small{H_2(t,\tau_j) = -\frac45\sum_{k=A,B}e^{-ikv_{eq}t+P_k\tau}\left\{\frac{P_k}{P_k-iR}-\frac{P_k}{P_k-iR+q_j}-\frac{2q_jP_k[1-e^{(iR-P_k-q_j)\tau}]}{P_k-iR+q_j}\right\},}\\
&\small{H_3(t,\tau_j)=\frac{\lambda}{\sqrt2\pi}e^{-ikv_{eq}t}\int_0^\infty\left\{\frac{1}{u+iR}-\frac{1}{u+iR-q_j}+\frac{2q_j[1-e^{(iR+u-q_j)\tau}]}{[u+iR-q_j]^2}\right\}}\nonumber\\
&\qquad \small{\times\frac{u^\frac54}{1-\sqrt2\lambda u^\frac54+\lambda^2u^\frac52}e^{-u\tau}d\tau.}
\end{align}
Here $j=0, 1$, $\tau_1=\tau_T$, $q_0=\tau_R^{\frac35}\tau_A^{\frac25}/\tau_0$, and $q_1=\tau_R^{\frac35}\tau_A^{\frac25}/\tau_T$. 

Fig. $5$ in ~\cite{beidler18a} shows that the forced island evolution is dramatically influenced by the amplitude of $\delta_1$ as well as $\Delta t_T$. As mentioned by Comisso et al ~\cite{comisso15b}, three possible nonlinear scenarios may occur. When the driven island width exceeds the resistive layer width, the Rutherford evolution takes place if the constant-$\psi$ is valid in the nonlinear regime. The Sweet-Parker evolution or even the plasmoid phase may occur in the nonlinear regime if the island does not satisfy the constant-$\psi$ assumption. Depending on the values $\delta_0$, $\delta_1$ and $\Delta t_T$, the forced island can nonlinearly evolve to the Rutherford regime, the Sweet-Parker regime, or the plasmoid regime.

\section{Quasi-linear forces and plasma response model}
Based on the linear solution for plasma response in the resistive-inertial regime, we derive the quasi-linear Maxwell (or, electromagnetic) force as well as the Reynolds force. The quasi-linear forces can influence the dynamics of plasma flow, which in turn affects the plasma response itself. Thus the formulations we developed for the quasi-linear forces are further used to construct the nonlinear plasma response model.
\subsection{Electromagnetic and Reynolds forces}
 From the reduced MHD model, we re-write the surface averaged poloidal momentum equation as
\begin{align}
\rho \partial_t \bar u_y &=  -\left \langle \partial_{xx}\psi_1 \partial_y \psi_1 \right \rangle/\mu_0 + \rho \left \langle \partial_{xx}\phi_1 \partial_y \phi_1 \right \rangle\nonumber\\
&= M(x) + R(x),
\end{align}
where $L_y = 2\pi/k$ and $\bar f = \left \langle f \right \rangle = \int_0^{L_y}f(x,y)/L_y$. $M(x)$ and $R(x)$ are Maxwell stress and Reynolds stress, respectively.

 Now let's focus on the Maxwell force. From the error field theory~\cite{fitz93, cole15a}, the Maxwell force can be expressed as
\begin{align}
F_m = \int M(x) dx = -\frac{k}{2\mu_0}[\Imag{\left(\psi_1^*\psi_1'\right)}]_{x = 0},
\end{align}
where $[f]_{x = 0}$ is the jump in $f$ across the resistive layer around $x=0$. Following the constant-$\psi$ assumption, the general form of the Maxwell force can be written as
\begin{align}
F_m = -\frac{k}{2\mu_0}\Imag\left(\psi_1^*(0)[\psi_1']_{x = 0}\right) = -\frac{k}{2\mu_0}|\psi_s|^2\Imag\Delta' = -\frac{k}{2\mu_0}E_{sc}|\psi_s||\psi_c| \sin{\varphi}\label{MXWTQ},
\end{align}
where $\Delta' = \frac{[\psi_1']}{\psi_1}|_{x = 0} = \Delta_0' + \Delta_e'\frac{\psi_c}{\psi_s}$, $\Imag\Delta' = \Delta'_{e}\Imag\left\{\frac{\psi_c}{\psi_s}\right\}$, $\psi_s = |\psi_s|e^{-i\varphi_s}$, $\psi_c = B_0|\delta_{\rm RMP}| e^{-i\varphi_c}$, and $\varphi = \varphi_s - \varphi_c$, $\varphi_c = \Omega t$. In our equilibrium, $E_{sc} = \Delta_e'=\frac{k}{\sinh{(ka)}}$ and $\Delta_0' = \frac{-k}{\tanh{(ka)}}$. The above general formula of electromagnetic force in Eq. \eqref{MXWTQ} has been often used to construct the nonlinear error field model~\cite{huang15a}.

\begin{figure}[htbp]
\setlength{\unitlength}{0.5cm}
\begin{center}
\includegraphics[width= 10 cm]{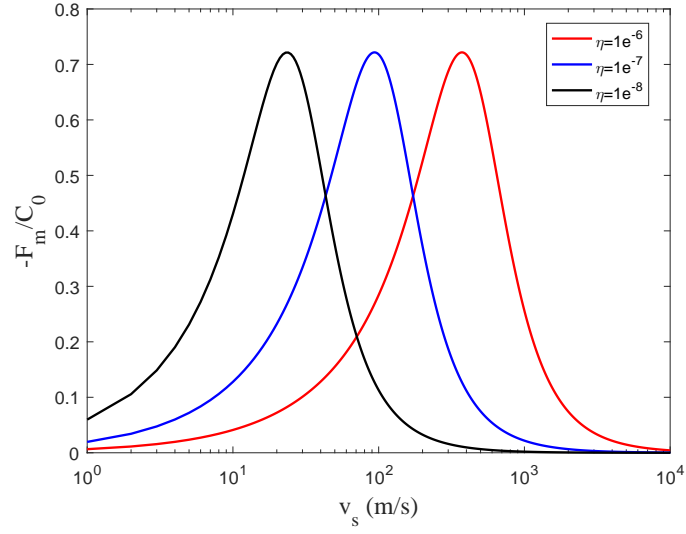}
\caption{Dependence of Maxwell force in steady state on plasma flow for different plasma resistivities. The red, blue and black curves represent $\eta=10^{-6}\Omega m$, $\eta=10^{-7}\Omega m$ , and $\eta=10^{-8}\Omega m$, respectively.
}
\label{fig7}
\end{center}
\end{figure}

When the island width is much less than the resistive layer width, the island evolution can be described by the island solution in Eq. \eqref{newHK} in the resistive-inertial regime. To derive the Maxwell force in such a regime, i.e. $W \ll \delta_{\rm layer}$, we substitute in Eq. \eqref{newHK} into Eq. \eqref{MXWTQ} and neglect the time-dependence of $\psi_s$, we obtain the Maxwell force in steady state as

\begin{align}
F_m &= -\frac{k}{2\mu_0}|\psi_s|^2\frac{k}{\tanh{(ka)}}\lambda|R_0|^{\frac54}\sgn{(\omega_0)}\sin{\frac58\pi}\nonumber\\
    &= -\frac{k\rho^\frac12}{2\mu_0^{\frac14}\eta^{\frac34}\alpha^\frac12}|\psi_s|^2\Delta_s^{ri}\sgn{(\omega_0)}|\omega_0|^{\frac54}\sin{\frac58\pi},
\end{align}
where $\alpha = \frac{kB_0}{a}$, $\Delta_s^{ri} = \frac{\lambda (ka)^{\frac32}}{\tanh{(ka)}}=\frac{3}{2\sqrt{2}}\approx 1$, $\omega_0=kv_{s}$, and $R_0= \omega_0 \tau_R^{\frac35} \tau_A^{\frac25}$. Here $v_s = v_0(x=0, t \rightarrow \infty)$ is the steady plasma flow at the rational surface. The parameter scaling of $F_m$ in resistive-inertial regime is consistent with the corresponding result in Ref. \onlinecite{cole15a}. In the limit $t \rightarrow \infty$, we re-write the steady Maxwell force as
\begin{align}
&F_m = -C_0 \frac{\lambda|R_0|^{\frac54}\sgn{(\omega_0)}\sin{\frac58\pi}}{1+\lambda^2|R_0|^{\frac52}+2\lambda|R_0|^{\frac54}\cos{\frac58\pi}}\label{FMAXWELL1},\\
&C_0 = \frac{k}{2\mu_0}[\frac{B_0\delta_{\rm RMP}}{\cosh{(ka)}}]^2\frac{k}{\tanh{(ka)}}\label{FMAXWELL2}.
\end{align}
$F_m \approx C_0 \lambda|R_0|^{\frac54}\sgn{(\omega_0)}\sin{\frac58\pi}$ when $\lambda|R_0|^{\frac54} \ll 1$, and as a result, the Maxwell force is proportional to $|\omega_0|^{\frac54}$. While $\lambda|R_0|^{\frac54} \gg 1$, $F_m \approx C_0 \sgn{(\omega_0)}\sin{\frac58\pi}/[\lambda|R_0|^{\frac54}]$, and the Maxwell force is proportional to $1/|\omega_0|^{\frac54}$ due to the flow shielding effect. As shown in Fig. \ref{fig7}, the Maxwell force from Eqs. \eqref{FMAXWELL1} and \eqref{FMAXWELL2} increases with the plasma flow at first. After the rigid flow exceeds a threshold, the force strongly decreases with $v_{s}$. In addition, the critical value increases with plasma resistivity. The above features of the Maxwell force are in good agreement with previous linear calculation results in Fig. 4(a) of ~\cite{furukawa09a}.

Similar to the Maxwell force, the Reynolds force can be expressed as
\begin{align}
F_r = \int R(x) dx = \frac{k\rho}{2}[\Imag\left(\phi_1^*\phi_1'\right)]_{x = 0}\label{ReydsF}.
\end{align}
From the definitions in Sec. III, for the steady state, we arrive at
\begin{align}
&\phi_1 = -\nu_sU_s = -\nu_s\sum a_n\hat D_n(\theta_s),\\
&a_n = -\frac{\int \theta_s\hat D_nd\theta_s}{n+\frac12}\frac{k}{B_0}\psi_s,\\
&\psi_s = \frac{B_0\delta_{\rm RMP}}{\cosh{(ka)}[1+\lambda(iR_0)^{\frac54}]},
\end{align}
where $\epsilon_s^4 = \frac{i\omega_0\tau_A^2}{4(ka)^2\tau_R}$, $\nu_s = \frac{\omega_0}{4\epsilon_sk^2}$, and $\theta_s = x/\epsilon_sa$. Together with Eq. ($41$) in ~\cite{furth63a} and the definition of Reynolds stress, the Reynolds force can be expressed as
\begin{align}
F_r = \int R(x) dx = -\frac{k}{2}\rho C_1|\nu_s|^2(\frac{k}{B_0})^2|\psi_s|^2\Imag{\frac{1}{\epsilon_sa}}=\frac{k}{2}\rho C_1|\nu_s|^2(\frac{k}{B_0})^2|\psi_s|^2|\frac{1}{\epsilon_sa}|\sgn{(\omega_0)}\sin{\frac{\pi}{8}},
\end{align}
where
\begin{align}
C_1 = -2^{\frac72}\sum_{l=0}^\infty\frac{l+\frac12}{(2l-1)(2l+\frac32)(2l+\frac72)}\frac{\Gamma(l+\frac12)}{\Gamma(l+1)} \approx 0.27 \times 2^{\frac72}.
\end{align}
More explicitly, we rewrite the steady state Maxwell force as well as the steady Reynolds force induced by static RMP as
\begin{align}
&F_m = -\frac{k}{2\mu_0}\frac{\lambda k}{\tanh{(ka)}}|\omega_0|^{\frac54}\tau_R^{\frac34}\tau_A^{\frac12}|\psi_s|^2\sgn{(\omega_0)}\sin{\frac58\pi},\\
&F_r = \frac{k}{2}\rho C_1\frac{[4(ka)^2]^{\frac34}}{16a(kB_0)^2\tau_A^2}|\omega_0|^{\frac54}\tau_R^{\frac34}\tau_A^{\frac12}|\psi_s|^2\sgn{(\omega_0)}\sin{\frac18\pi},
\end{align}
which leads to
\begin{align}
&F_r = -C_2F_m,\label{C2}\\
&C_2 = \frac{\sqrt{2}C_1}{8(ka)^{\frac32}}\frac{\tanh{(ka)}}{\lambda}\frac{\sin{\frac{\pi}{8}}}{\sin{\frac{5\pi}{8}}}=\frac{C_1}{6}\frac{\sin{\frac{\pi}{8}}}{\sin{\frac{5\pi}{8}}} \approx 0.3.
\end{align}
Obviously, the Reynolds force $F_r$ is opposite sign to the Maxwell force $F_m$ and $F_r\sim -0.3F_m \ll F_m$.

For comparison, previous study relates $\phi_1$ to $\psi_1$ in the outer region as in $\phi_1 = \frac{\omega_0}{\alpha x}\psi_1$~\cite{cole15a}. By substituting such a relationship and evaluating the jump condition at $x\sim\pm\delta_{\rm layer}$, they arrive at
\begin{align}
F_r \sim (\frac{\omega_0}{\alpha|\delta_{\rm layer}|})^2(-F_m) \label{coleFr},
\end{align}
where
\begin{align}
\delta_{\rm layer} = [\frac{\rho|\omega_0|\eta}{\alpha^2\mu_0}]^{\frac14}e^{i\frac{\pi}{8} \sgn{(\omega_0)}}.
\end{align}
Similarly, they find that the Reynolds force is much less than the Maxwell force by the factor $\rho\mu_0(\frac{\omega_0}{\alpha|\delta_{\rm layer}|})^2=\frac{1}{(ka)^2}\omega_0^{\frac32}\tau_A\tau_R^{\frac12}$, in contrast to the factor $C_2$ in Eq. \eqref{C2} found in our study.

The parameter scalings of $F_r$, as well as the relationship between $F_r$ and $F_m$, are quite different from the previous result in ~\cite{cole15a}. In fact, the parameter scaling of $F_r$ is the same with $F_m$ in our analytical result. In other words, the ratio between the two steady state forces in the resistive-inertial regime is independent of equilibrium parameters in presence of uniform plasma flow.


\subsection{Nonlinear plasma response model in absence of no-slip condition}

The above quasi-linear forces can be used to construct nonlinear model for plasma response and flow evolution. When the island width is much less than the resistive layer width $\delta_{\rm layer}$, the island evolution can be described by the linear solution of plasma response in the resistive-inertial regime as follows
\begin{align}
&\psi_s= \frac{B_0}{\cosh{(ka)}}e^{-i\varphi_{\rm temp}(t)}\int_0^t G_2(t-t')\delta_{\rm RMP}(t')e^{i\varphi_{\rm temp}(t')}dt',\\
&F_m= -\frac{k}{2\mu_0}E_{sc}|\psi_s|^2\Imag\left\{\frac{\psi_c}{\psi_s}\right\},\\
&\rho_0\frac{\partial \delta v_0}{\partial t} = F_m\delta(x) + \nu_{\perp}\frac{\partial^2\delta v_0}{\partial x^2}\label{ERMod1},
\end{align}
where $\delta(x)$ is the Dirac $\delta$ function. To obtain the above analytical model, only $W \ll \delta_{\rm layer}$ and constant-$\psi$ are assumed. Note that the viscous term is added in Eq. \eqref{ERMod1} to balance the electromagnetic force although we neglect such an effect in linear island solution. This maybe reasonable since we assume that $P_r\ll 1$, and we will extend our theory to include the viscous effect in the future.


When the island width exceeds the resistive layer width $\delta_{\rm layer}$, say, $\delta_{\rm layer}\ll W\ll a$, and the constant-$\psi$ assumption is still valid, the driven island enters the nonlinear Rutherford regime. In the standard error field theory, the no-slip condition for the phase equation is widely used. However, as we discussed in Sec. III, the no-slip condition is not satisfied in the linear constant-$\psi$ resistive-inertial regime. Furthermore, the no-slip condition would require the plasma be static on rational surface in the full RMP penetration state, which does not agree with previous nonlinear reduced MHD simulations~\cite{yu08b}. Extending Rutherford's original work~\cite{rutherford73}, we derive a set of island width and phase equations without the constraint of no-slip condition. Following the method in Ref. \onlinecite{rutherford73}, we arrive at the asymptotic matching relation
\begin{align}
\Delta'|\psi_s|^{\frac12}e^{-i\varphi_s} = \frac{4 A \mu_0}{\eta(2B_0/a)^{\frac12}}[\frac{d}{dt}+i\omega_s](|\psi_s|e^{-i\varphi_s})\label{rutherford},
\end{align}
where $A \approx 0.7$ and $\omega_s=kv_0(x=0)$ is the plasma angular rotation frequency at the rational surface. From the real part and the imaginary part, we obtain the island width and phase equations, respectively. On the other hand, the plasma flow can be modified by the Maxwell and viscous forces induced by plasma response. To close the system, the nonlinear force balance equation is needed. Here we write the nonlinear plasma response model as
\begin{align}
&\frac{2A\tau_R}{\sqrt2 a^2}\frac{dW}{dt} = \Delta'_0 + \Delta'_{e}\frac{W_c^2}{W^2}\cos{\varphi}\label{WEQ},\\
&\frac{d\varphi}{dt} = \omega_s + \frac{\sqrt2 a^2}{2A\tau_R}\Delta'_{e}\frac{W_c^2}{W^3}\sin{\varphi}\label{phaseEQ},\\
&\rho_0\frac{\partial \delta v_0}{\partial t} = -\frac{k}{2\mu_0}E_{sc}|\psi_s||\psi_c| \sin{\varphi}\delta(x) + \nu_{\perp}\frac{\partial^2\delta v_0}{\partial x^2},
\end{align}
where $W=2\sqrt{\frac{a}{B_0}|\psi_s|}$ is the island width and $W_c = 2\sqrt{(\frac{a}{B_0})|\psi_c|}$. The no-slip condition assumed in the conventional error field theory is not imposed in the above nonlinear plasma response model. From the phase equation Eq.\eqref{phaseEQ}, one finds that the island oscillation frequency depends on but does not necessarily equal to the plasma flow frequency at the rational surface. For simplicity, here we only derive the linear drive and external field terms in Rutherford equation in Eq.\eqref{WEQ}. Note that the Rutherford equation is a nonlinear equation even in absence of the nonlinear saturation term. Similarly, the nonlinear terms in phase equation also is not included here either. We also note that Eq. \eqref{rutherford} is obtained by neglecting the inertial and viscous effects. Such effects as well as the nonlinear saturation terms should be included in the nonlinear response model in the future.

\section{Summary and discussion}
In summary, we have developed a new theory model for the nonlinear plasma response to external magnetic perturbation in absence of the no-slip condition. The model is composed of the equations for the evolution of both width and phase of magnetic island due to forced reconnection driven by the external magnetic perturbation, and the force-balance equation for the plasma flow. When the island width is much less than the resistive layer width, the island growth is governed by the linear Hahm-Kulsrud-Taylor solution in presence of time-dependent plasma flow. Based on the standard asymptotic matching and Laplace transform, we have extended the linear response solution to include the equilibrium flow in both the inertial and the resistive-inertial regimes. In particular, the plasma flow in our new island solution in the resistive-inertial regime can be time-dependent. The island solution is used to construct the quasi-linear electromagnetic force, which together with viscous force, contributes to the driving and damping of plasma flow. In case of uniform flow, the ratio of corresponding Maxwell and Reynolds forces in steady state proves to be a constant independent of equilibrium, which is about $3$. When the island width is much larger than the resistive layer width, the evolution of both island width and phase can be described using the newly developed nonlinear model. The no-slip condition assumed in the conventional error field theory is not imposed here, where the island oscillation frequency depends on but does not necessarily equal to the plasma flow frequency at the rational surface. Using the new developed plasma response model, recent simulation results can be better understood~\cite{becoulet12a, furukawa09a, yu08b}.

The theory in this work is for the forced island solution in both resistive-inertial and Rutherford regimes. It is our first step towards constructing a general plasma response model in absence of no-slip condition. Several physics elements of island-flow interaction are missing in the developed model that may potentially have significant impacts, but they are well beyond the scope of this report. For example, viscosity, two-fluid, and finite-Larmor-radius effects are known to have strong influence over plasma response to RMPs near resonant surfaces~\cite{fitz98a, wael12}. Furthermore, the plasma response model here is developed in the two limits of $W \ll \delta_{\rm layer}$ and $W \gg \delta_{\rm layer}$. The more complete model connecting the two regimes yet to be built. We plan to address these important issues in future studies.

\begin{acknowledgments}
This work was supported by the Fundamental Research Funds for the Central Universities at Huazhong University of Science and Technology Grant No. 2019kfyXJJS193, the National Natural Science Foundation of China Grant No. 11775221, the National Magnetic Confinement Fusion Science Program of China Grant No. 2015GB101004, the Young Elite Scientists Sponsorship Program by CAST Grant No. 2017QNRC001, and U.S. Department of Energy Grant Nos. DE-FG02-86ER53218 and DE-SC0018001.
\end{acknowledgments}

\newpage

\newpage

\newpage
\newpage

\newpage

\newpage


\begin{thebibliography}{39}
\expandafter\ifx\csname natexlab\endcsname\relax\def\natexlab#1{#1}\fi
\expandafter\ifx\csname bibnamefont\endcsname\relax
  \def\bibnamefont#1{#1}\fi
\expandafter\ifx\csname bibfnamefont\endcsname\relax
  \def\bibfnamefont#1{#1}\fi
\expandafter\ifx\csname citenamefont\endcsname\relax
  \def\citenamefont#1{#1}\fi
\expandafter\ifx\csname url\endcsname\relax
  \def\url#1{\texttt{#1}}\fi
\expandafter\ifx\csname urlprefix\endcsname\relax\def\urlprefix{URL }\fi
\providecommand{\bibinfo}[2]{#2}
\providecommand{\eprint}[2][]{\url{#2}}

\bibitem[{\citenamefont{Rao et~al.}(2013)\citenamefont{Rao, Ding, Yu, Jin, Hu,
  Yi, Nan, Wang, Zhang, and Zhuang}}]{rao13a}
\bibinfo{author}{\bibfnamefont{B.}~\bibnamefont{Rao}},
  \bibinfo{author}{\bibfnamefont{Y.~H.} \bibnamefont{Ding}},
  \bibinfo{author}{\bibfnamefont{K.~X.} \bibnamefont{Yu}},
  \bibinfo{author}{\bibfnamefont{W.}~\bibnamefont{Jin}},
  \bibinfo{author}{\bibfnamefont{Q.~M.} \bibnamefont{Hu}},
  \bibinfo{author}{\bibfnamefont{B.}~\bibnamefont{Yi}},
  \bibinfo{author}{\bibfnamefont{J.~Y.} \bibnamefont{Nan}},
  \bibinfo{author}{\bibfnamefont{N.~C.} \bibnamefont{Wang}},
  \bibinfo{author}{\bibfnamefont{M.}~\bibnamefont{Zhang}}, \bibnamefont{and}
  \bibinfo{author}{\bibfnamefont{G.}~\bibnamefont{Zhuang}},
  \bibinfo{journal}{Review of Scientific Instruments}
  \textbf{\bibinfo{volume}{84}}, \bibinfo{pages}{043504}
  (\bibinfo{year}{2013}).

\bibitem[{\citenamefont{Wang et~al.}(2016)\citenamefont{Wang, Sun, Qian, Shi,
  Shen, Gu, Liu, Guo, Chu, He et~al.}}]{wang16a}
\bibinfo{author}{\bibfnamefont{H.-H.} \bibnamefont{Wang}},
  \bibinfo{author}{\bibfnamefont{Y.-W.} \bibnamefont{Sun}},
  \bibinfo{author}{\bibfnamefont{J.-P.} \bibnamefont{Qian}},
  \bibinfo{author}{\bibfnamefont{T.-H.} \bibnamefont{Shi}},
  \bibinfo{author}{\bibfnamefont{B.}~\bibnamefont{Shen}},
  \bibinfo{author}{\bibfnamefont{S.}~\bibnamefont{Gu}},
  \bibinfo{author}{\bibfnamefont{Y.-Q.} \bibnamefont{Liu}},
  \bibinfo{author}{\bibfnamefont{W.-F.} \bibnamefont{Guo}},
  \bibinfo{author}{\bibfnamefont{N.}~\bibnamefont{Chu}},
  \bibinfo{author}{\bibfnamefont{K.-Y.} \bibnamefont{He}},
  \bibnamefont{et~al.}, \bibinfo{journal}{Nuclear Fusion}
  \textbf{\bibinfo{volume}{56}}, \bibinfo{pages}{066011}
  (\bibinfo{year}{2016}).

\bibitem[{\citenamefont{Nave and Wesson}(1990)}]{nave90}
\bibinfo{author}{\bibfnamefont{M.}~\bibnamefont{Nave}} \bibnamefont{and}
  \bibinfo{author}{\bibfnamefont{J.}~\bibnamefont{Wesson}},
  \bibinfo{journal}{Nuclear Fusion} \textbf{\bibinfo{volume}{30}},
  \bibinfo{pages}{2575} (\bibinfo{year}{1990}).

\bibitem[{\citenamefont{Liu et~al.}(2010)\citenamefont{Liu, Chu, In, and
  Okabayashi}}]{liu10a}
\bibinfo{author}{\bibfnamefont{Y.}~\bibnamefont{Liu}},
  \bibinfo{author}{\bibfnamefont{M.~S.} \bibnamefont{Chu}},
  \bibinfo{author}{\bibfnamefont{Y.}~\bibnamefont{In}}, \bibnamefont{and}
  \bibinfo{author}{\bibfnamefont{M.}~\bibnamefont{Okabayashi}},
  \bibinfo{journal}{Physics of Plasmas} \textbf{\bibinfo{volume}{17}},
  \bibinfo{pages}{072510} (\bibinfo{year}{2010}).

\bibitem[{\citenamefont{Evans et~al.}(2005)\citenamefont{Evans, Moyer, Watkins,
  Osborne, Thomas, Becoulet, Boedo, Doyle, Fenstermacher, Finken
  et~al.}}]{evans05a}
\bibinfo{author}{\bibfnamefont{T.}~\bibnamefont{Evans}},
  \bibinfo{author}{\bibfnamefont{R.}~\bibnamefont{Moyer}},
  \bibinfo{author}{\bibfnamefont{J.}~\bibnamefont{Watkins}},
  \bibinfo{author}{\bibfnamefont{T.}~\bibnamefont{Osborne}},
  \bibinfo{author}{\bibfnamefont{P.}~\bibnamefont{Thomas}},
  \bibinfo{author}{\bibfnamefont{M.}~\bibnamefont{Becoulet}},
  \bibinfo{author}{\bibfnamefont{J.}~\bibnamefont{Boedo}},
  \bibinfo{author}{\bibfnamefont{E.}~\bibnamefont{Doyle}},
  \bibinfo{author}{\bibfnamefont{M.}~\bibnamefont{Fenstermacher}},
  \bibinfo{author}{\bibfnamefont{K.}~\bibnamefont{Finken}},
  \bibnamefont{et~al.}, \bibinfo{journal}{Nuclear Fusion}
  \textbf{\bibinfo{volume}{45}}, \bibinfo{pages}{595} (\bibinfo{year}{2005}).

\bibitem[{\citenamefont{Suttrop et~al.}(2011)\citenamefont{Suttrop, Eich,
  Fuchs, G\"unter, Janzer, Herrmann, Kallenbach, Lang, Lunt, Maraschek
  et~al.}}]{sut11a}
\bibinfo{author}{\bibfnamefont{W.}~\bibnamefont{Suttrop}},
  \bibinfo{author}{\bibfnamefont{T.}~\bibnamefont{Eich}},
  \bibinfo{author}{\bibfnamefont{J.~C.} \bibnamefont{Fuchs}},
  \bibinfo{author}{\bibfnamefont{S.}~\bibnamefont{G\"unter}},
  \bibinfo{author}{\bibfnamefont{A.}~\bibnamefont{Janzer}},
  \bibinfo{author}{\bibfnamefont{A.}~\bibnamefont{Herrmann}},
  \bibinfo{author}{\bibfnamefont{A.}~\bibnamefont{Kallenbach}},
  \bibinfo{author}{\bibfnamefont{P.~T.} \bibnamefont{Lang}},
  \bibinfo{author}{\bibfnamefont{T.}~\bibnamefont{Lunt}},
  \bibinfo{author}{\bibfnamefont{M.}~\bibnamefont{Maraschek}},
  \bibnamefont{et~al.}, \bibinfo{journal}{Phys. Rev. Lett.}
  \textbf{\bibinfo{volume}{106}}, \bibinfo{pages}{225004}
  (\bibinfo{year}{2011}).

\bibitem[{\citenamefont{Hu et~al.}(2015)\citenamefont{Hu, Wang, Yu, Ding, Rao,
  Chen, and Jin}}]{hu15a}
\bibinfo{author}{\bibfnamefont{Q.}~\bibnamefont{Hu}},
  \bibinfo{author}{\bibfnamefont{N.}~\bibnamefont{Wang}},
  \bibinfo{author}{\bibfnamefont{Q.}~\bibnamefont{Yu}},
  \bibinfo{author}{\bibfnamefont{Y.}~\bibnamefont{Ding}},
  \bibinfo{author}{\bibfnamefont{B.}~\bibnamefont{Rao}},
  \bibinfo{author}{\bibfnamefont{Z.}~\bibnamefont{Chen}}, \bibnamefont{and}
  \bibinfo{author}{\bibfnamefont{H.}~\bibnamefont{Jin}},
  \bibinfo{journal}{Plasma Physics and Controlled Fusion}
  \textbf{\bibinfo{volume}{58}}, \bibinfo{pages}{025001}
  (\bibinfo{year}{2015}).

\bibitem[{\citenamefont{Ding et~al.}(2018)\citenamefont{Ding, Chen, Chen, Yang,
  Wang, Hu, Rao, Chen, Cheng, Gao et~al.}}]{ding2018}
\bibinfo{author}{\bibfnamefont{Y.}~\bibnamefont{Ding}},
  \bibinfo{author}{\bibfnamefont{Z.}~\bibnamefont{Chen}},
  \bibinfo{author}{\bibfnamefont{Z.}~\bibnamefont{Chen}},
  \bibinfo{author}{\bibfnamefont{Z.}~\bibnamefont{Yang}},
  \bibinfo{author}{\bibfnamefont{N.}~\bibnamefont{Wang}},
  \bibinfo{author}{\bibfnamefont{Q.}~\bibnamefont{Hu}},
  \bibinfo{author}{\bibfnamefont{B.}~\bibnamefont{Rao}},
  \bibinfo{author}{\bibfnamefont{J.}~\bibnamefont{Chen}},
  \bibinfo{author}{\bibfnamefont{Z.}~\bibnamefont{Cheng}},
  \bibinfo{author}{\bibfnamefont{L.}~\bibnamefont{Gao}}, \bibnamefont{et~al.},
  \bibinfo{journal}{Plasma Science and Technology}
  \textbf{\bibinfo{volume}{20}}, \bibinfo{pages}{125101}
  (\bibinfo{year}{2018}).

\bibitem[{\citenamefont{Liang et~al.}(2019)\citenamefont{Liang, Wang, Ding,
  Chen, Chen, Yang, Hu, Cheng, Wang, Jiang et~al.}}]{liang2019}
\bibinfo{author}{\bibfnamefont{Y.}~\bibnamefont{Liang}},
  \bibinfo{author}{\bibfnamefont{N.}~\bibnamefont{Wang}},
  \bibinfo{author}{\bibfnamefont{Y.}~\bibnamefont{Ding}},
  \bibinfo{author}{\bibfnamefont{Z.}~\bibnamefont{Chen}},
  \bibinfo{author}{\bibfnamefont{Z.}~\bibnamefont{Chen}},
  \bibinfo{author}{\bibfnamefont{Z.}~\bibnamefont{Yang}},
  \bibinfo{author}{\bibfnamefont{Q.}~\bibnamefont{Hu}},
  \bibinfo{author}{\bibfnamefont{Z.}~\bibnamefont{Cheng}},
  \bibinfo{author}{\bibfnamefont{L.}~\bibnamefont{Wang}},
  \bibinfo{author}{\bibfnamefont{Z.}~\bibnamefont{Jiang}},
  \bibnamefont{et~al.}, \bibinfo{journal}{Nuclear Fusion}
  \textbf{\bibinfo{volume}{59}}, \bibinfo{pages}{112016}
  (\bibinfo{year}{2019}).

\bibitem[{\citenamefont{Hu et~al.}(2012)\citenamefont{Hu, Yu, Rao, Ding, Hu,
  Zhuang, and the J-TEXT~Team}}]{hu12a}
\bibinfo{author}{\bibfnamefont{Q.}~\bibnamefont{Hu}},
  \bibinfo{author}{\bibfnamefont{Q.}~\bibnamefont{Yu}},
  \bibinfo{author}{\bibfnamefont{B.}~\bibnamefont{Rao}},
  \bibinfo{author}{\bibfnamefont{Y.}~\bibnamefont{Ding}},
  \bibinfo{author}{\bibfnamefont{X.}~\bibnamefont{Hu}},
  \bibinfo{author}{\bibfnamefont{G.}~\bibnamefont{Zhuang}}, \bibnamefont{and}
  \bibinfo{author}{\bibnamefont{the J-TEXT~Team}}, \bibinfo{journal}{Nuclear
  Fusion} \textbf{\bibinfo{volume}{52}}, \bibinfo{pages}{083011}
  (\bibinfo{year}{2012}).

\bibitem[{\citenamefont{Hu et~al.}(2013)\citenamefont{Hu, Rao, Yu, Ding,
  Zhuang, Jin, and Hu}}]{hu13a}
\bibinfo{author}{\bibfnamefont{Q.}~\bibnamefont{Hu}},
  \bibinfo{author}{\bibfnamefont{B.}~\bibnamefont{Rao}},
  \bibinfo{author}{\bibfnamefont{Q.}~\bibnamefont{Yu}},
  \bibinfo{author}{\bibfnamefont{Y.}~\bibnamefont{Ding}},
  \bibinfo{author}{\bibfnamefont{G.}~\bibnamefont{Zhuang}},
  \bibinfo{author}{\bibfnamefont{W.}~\bibnamefont{Jin}}, \bibnamefont{and}
  \bibinfo{author}{\bibfnamefont{X.}~\bibnamefont{Hu}},
  \bibinfo{journal}{Physics of Plasmas} \textbf{\bibinfo{volume}{20}},
  \bibinfo{pages}{092502} (\bibinfo{year}{2013}).

\bibitem[{\citenamefont{Jin et~al.}(2015)\citenamefont{Jin, Hu, Wang, Rao,
  Ding, Li, Li, and Xie}}]{jin15}
\bibinfo{author}{\bibfnamefont{H.}~\bibnamefont{Jin}},
  \bibinfo{author}{\bibfnamefont{Q.}~\bibnamefont{Hu}},
  \bibinfo{author}{\bibfnamefont{N.}~\bibnamefont{Wang}},
  \bibinfo{author}{\bibfnamefont{B.}~\bibnamefont{Rao}},
  \bibinfo{author}{\bibfnamefont{Y.}~\bibnamefont{Ding}},
  \bibinfo{author}{\bibfnamefont{D.}~\bibnamefont{Li}},
  \bibinfo{author}{\bibfnamefont{M.}~\bibnamefont{Li}}, \bibnamefont{and}
  \bibinfo{author}{\bibfnamefont{S.}~\bibnamefont{Xie}},
  \bibinfo{journal}{Plasma Physics and Controlled Fusion}
  \textbf{\bibinfo{volume}{57}}, \bibinfo{pages}{104007}
  (\bibinfo{year}{2015}).

\bibitem[{\citenamefont{Hu and Yu}(2016)}]{hu16}
\bibinfo{author}{\bibfnamefont{Q.}~\bibnamefont{Hu}} \bibnamefont{and}
  \bibinfo{author}{\bibfnamefont{Q.}~\bibnamefont{Yu}},
  \bibinfo{journal}{Nuclear Fusion} \textbf{\bibinfo{volume}{56}},
  \bibinfo{pages}{034001} (\bibinfo{year}{2016}).

\bibitem[{\citenamefont{Fitzpatrick}(1993{\natexlab{a}})}]{fitz93}
\bibinfo{author}{\bibfnamefont{R.}~\bibnamefont{Fitzpatrick}},
  \bibinfo{journal}{Nuclear Fusion} \textbf{\bibinfo{volume}{33}},
  \bibinfo{pages}{1049} (\bibinfo{year}{1993}{\natexlab{a}}).

\bibitem[{\citenamefont{Fitzpatrick et~al.}(2001)\citenamefont{Fitzpatrick,
  Rossi, and Yu}}]{fitz01a}
\bibinfo{author}{\bibfnamefont{R.}~\bibnamefont{Fitzpatrick}},
  \bibinfo{author}{\bibfnamefont{E.}~\bibnamefont{Rossi}}, \bibnamefont{and}
  \bibinfo{author}{\bibfnamefont{E.~P.} \bibnamefont{Yu}},
  \bibinfo{journal}{Physics of Plasmas} \textbf{\bibinfo{volume}{8}},
  \bibinfo{pages}{4489} (\bibinfo{year}{2001}).

\bibitem[{\citenamefont{Fitzpatrick}(2012)}]{fitz12a}
\bibinfo{author}{\bibfnamefont{R.}~\bibnamefont{Fitzpatrick}},
  \bibinfo{journal}{Plasma Physics and Controlled Fusion}
  \textbf{\bibinfo{volume}{54}}, \bibinfo{pages}{094002}
  (\bibinfo{year}{2012}).

\bibitem[{\citenamefont{Fitzpatrick}(2018)}]{fitz18a}
\bibinfo{author}{\bibfnamefont{R.}~\bibnamefont{Fitzpatrick}},
  \bibinfo{journal}{Physics of Plasmas} \textbf{\bibinfo{volume}{25}},
  \bibinfo{pages}{082513} (\bibinfo{year}{2018}).

\bibitem[{\citenamefont{Huang and Zhu}(2015)}]{huang15a}
\bibinfo{author}{\bibfnamefont{W.}~\bibnamefont{Huang}} \bibnamefont{and}
  \bibinfo{author}{\bibfnamefont{P.}~\bibnamefont{Zhu}},
  \bibinfo{journal}{Physics of Plasmas} \textbf{\bibinfo{volume}{22}},
  \bibinfo{pages}{032502} (\bibinfo{year}{2015}).

\bibitem[{\citenamefont{Huang and Zhu}(2016)}]{huang16a}
\bibinfo{author}{\bibfnamefont{W.}~\bibnamefont{Huang}} \bibnamefont{and}
  \bibinfo{author}{\bibfnamefont{P.}~\bibnamefont{Zhu}},
  \bibinfo{journal}{Physics of Plasmas} \textbf{\bibinfo{volume}{23}},
  \bibinfo{pages}{032505} (\bibinfo{year}{2016}).

\bibitem[{\citenamefont{Yu et~al.}(2008)\citenamefont{Yu, Gunter, Kikuchi, and
  Finken}}]{yu08b}
\bibinfo{author}{\bibfnamefont{Q.}~\bibnamefont{Yu}},
  \bibinfo{author}{\bibfnamefont{S.}~\bibnamefont{Gunter}},
  \bibinfo{author}{\bibfnamefont{Y.}~\bibnamefont{Kikuchi}}, \bibnamefont{and}
  \bibinfo{author}{\bibfnamefont{K.}~\bibnamefont{Finken}},
  \bibinfo{journal}{Nuclear Fusion} \textbf{\bibinfo{volume}{48}},
  \bibinfo{pages}{024007} (\bibinfo{year}{2008}).

\bibitem[{\citenamefont{Hahm and Kulsrud}(1985)}]{hahm85}
\bibinfo{author}{\bibfnamefont{T.~S.} \bibnamefont{Hahm}} \bibnamefont{and}
  \bibinfo{author}{\bibfnamefont{R.~M.} \bibnamefont{Kulsrud}},
  \bibinfo{journal}{Physics of Fluids} \textbf{\bibinfo{volume}{28}},
  \bibinfo{pages}{2412} (\bibinfo{year}{1985}).

\bibitem[{\citenamefont{Fitzpatrick and Hender}(1991)}]{fitz91a}
\bibinfo{author}{\bibfnamefont{R.}~\bibnamefont{Fitzpatrick}} \bibnamefont{and}
  \bibinfo{author}{\bibfnamefont{T.~C.} \bibnamefont{Hender}},
  \bibinfo{journal}{Physics of Fluids B} \textbf{\bibinfo{volume}{3}},
  \bibinfo{pages}{644} (\bibinfo{year}{1991}).

\bibitem[{\citenamefont{Fitzpatrick}(1998)}]{fitz98a}
\bibinfo{author}{\bibfnamefont{R.}~\bibnamefont{Fitzpatrick}},
  \bibinfo{journal}{Physics of Plasmas} \textbf{\bibinfo{volume}{5}},
  \bibinfo{pages}{3325} (\bibinfo{year}{1998}).

\bibitem[{\citenamefont{Wang and Bhattacharjee}(1992)}]{wang92a}
\bibinfo{author}{\bibfnamefont{X.}~\bibnamefont{Wang}} \bibnamefont{and}
  \bibinfo{author}{\bibfnamefont{A.}~\bibnamefont{Bhattacharjee}},
  \bibinfo{journal}{Physics of Fluids B} \textbf{\bibinfo{volume}{4}},
  \bibinfo{pages}{1795} (\bibinfo{year}{1992}).

\bibitem[{\citenamefont{Fitzpatrick}(2003)}]{fitz03a}
\bibinfo{author}{\bibfnamefont{R.}~\bibnamefont{Fitzpatrick}},
  \bibinfo{journal}{Physics of Plasmas} \textbf{\bibinfo{volume}{10}},
  \bibinfo{pages}{2304} (\bibinfo{year}{2003}).

\bibitem[{\citenamefont{Fitzpatrick}(2004{\natexlab{a}})}]{fitz04a}
\bibinfo{author}{\bibfnamefont{R.}~\bibnamefont{Fitzpatrick}},
  \bibinfo{journal}{Physics of Plasmas} \textbf{\bibinfo{volume}{11}},
  \bibinfo{pages}{937} (\bibinfo{year}{2004}{\natexlab{a}}).

\bibitem[{\citenamefont{Fitzpatrick}(2004{\natexlab{b}})}]{fitz04b}
\bibinfo{author}{\bibfnamefont{R.}~\bibnamefont{Fitzpatrick}},
  \bibinfo{journal}{Physics of Plasmas} \textbf{\bibinfo{volume}{11}},
  \bibinfo{pages}{3961} (\bibinfo{year}{2004}{\natexlab{b}}).

\bibitem[{\citenamefont{Comisso et~al.}(2015)\citenamefont{Comisso, Grasso, and
  Waelbroeck}}]{comisso15b}
\bibinfo{author}{\bibfnamefont{L.}~\bibnamefont{Comisso}},
  \bibinfo{author}{\bibfnamefont{D.}~\bibnamefont{Grasso}}, \bibnamefont{and}
  \bibinfo{author}{\bibfnamefont{F.~L.} \bibnamefont{Waelbroeck}},
  \bibinfo{journal}{Physics of Plasmas} \textbf{\bibinfo{volume}{22}},
  \bibinfo{pages}{042109} (\bibinfo{year}{2015}).

\bibitem[{\citenamefont{Dewar et~al.}(2013)\citenamefont{Dewar, Bhattacharjee,
  Kulsrud, and Wright}}]{dewar13a}
\bibinfo{author}{\bibfnamefont{R.~L.} \bibnamefont{Dewar}},
  \bibinfo{author}{\bibfnamefont{A.}~\bibnamefont{Bhattacharjee}},
  \bibinfo{author}{\bibfnamefont{R.~M.} \bibnamefont{Kulsrud}},
  \bibnamefont{and} \bibinfo{author}{\bibfnamefont{A.~M.}
  \bibnamefont{Wright}}, \bibinfo{journal}{Physics of Plasmas}
  \textbf{\bibinfo{volume}{20}}, \bibinfo{pages}{082103}
  (\bibinfo{year}{2013}).

\bibitem[{\citenamefont{Vekstein and Kusano}(2015)}]{vek15a}
\bibinfo{author}{\bibfnamefont{G.}~\bibnamefont{Vekstein}} \bibnamefont{and}
  \bibinfo{author}{\bibfnamefont{K.}~\bibnamefont{Kusano}},
  \bibinfo{journal}{Physics of Plasmas} \textbf{\bibinfo{volume}{22}},
  \bibinfo{pages}{090707} (\bibinfo{year}{2015}).

\bibitem[{\citenamefont{Cole et~al.}(2015)\citenamefont{Cole, Finn, Hegna, and
  Terry}}]{cole15a}
\bibinfo{author}{\bibfnamefont{A.~J.} \bibnamefont{Cole}},
  \bibinfo{author}{\bibfnamefont{J.~M.} \bibnamefont{Finn}},
  \bibinfo{author}{\bibfnamefont{C.~C.} \bibnamefont{Hegna}}, \bibnamefont{and}
  \bibinfo{author}{\bibfnamefont{P.~W.} \bibnamefont{Terry}},
  \bibinfo{journal}{Physics of Plasmas} \textbf{\bibinfo{volume}{22}},
  \bibinfo{pages}{102514} (\bibinfo{year}{2015}).

\bibitem[{\citenamefont{Beidler et~al.}(2017)\citenamefont{Beidler, Callen,
  Hegna, and Sovinec}}]{beidler17a}
\bibinfo{author}{\bibfnamefont{M.~T.} \bibnamefont{Beidler}},
  \bibinfo{author}{\bibfnamefont{J.~D.} \bibnamefont{Callen}},
  \bibinfo{author}{\bibfnamefont{C.~C.} \bibnamefont{Hegna}}, \bibnamefont{and}
  \bibinfo{author}{\bibfnamefont{C.~R.} \bibnamefont{Sovinec}},
  \bibinfo{journal}{Physics of Plasmas} \textbf{\bibinfo{volume}{24}},
  \bibinfo{pages}{052508} (\bibinfo{year}{2017}).

\bibitem[{\citenamefont{Beidler et~al.}(2018)\citenamefont{Beidler, Callen,
  Hegna, and Sovinec}}]{beidler18a}
\bibinfo{author}{\bibfnamefont{M.~T.} \bibnamefont{Beidler}},
  \bibinfo{author}{\bibfnamefont{J.~D.} \bibnamefont{Callen}},
  \bibinfo{author}{\bibfnamefont{C.~C.} \bibnamefont{Hegna}}, \bibnamefont{and}
  \bibinfo{author}{\bibfnamefont{C.~R.} \bibnamefont{Sovinec}},
  \bibinfo{journal}{Physics of Plasmas} \textbf{\bibinfo{volume}{25}},
  \bibinfo{pages}{082507} (\bibinfo{year}{2018}).

\bibitem[{\citenamefont{Fitzpatrick}(1993{\natexlab{b}})}]{fitz93a}
\bibinfo{author}{\bibfnamefont{R.}~\bibnamefont{Fitzpatrick}},
  \bibinfo{journal}{Nuclear Fusion} \textbf{\bibinfo{volume}{33}},
  \bibinfo{pages}{1049} (\bibinfo{year}{1993}{\natexlab{b}}).

\bibitem[{\citenamefont{Furth et~al.}(1963)\citenamefont{Furth, Killeen, and
  Rosenbluth}}]{furth63a}
\bibinfo{author}{\bibfnamefont{H.~P.} \bibnamefont{Furth}},
  \bibinfo{author}{\bibfnamefont{J.}~\bibnamefont{Killeen}}, \bibnamefont{and}
  \bibinfo{author}{\bibfnamefont{M.~N.} \bibnamefont{Rosenbluth}},
  \bibinfo{journal}{Physics of Fluids} \textbf{\bibinfo{volume}{6}},
  \bibinfo{pages}{459} (\bibinfo{year}{1963}).

\bibitem[{\citenamefont{Becoulet et~al.}(2012)\citenamefont{Becoulet, Orain,
  Maget, Mellet, Garbet, Nardon, Huysmans, Casper, Loarte, Cahyna
  et~al.}}]{becoulet12a}
\bibinfo{author}{\bibfnamefont{M.}~\bibnamefont{Becoulet}},
  \bibinfo{author}{\bibfnamefont{F.}~\bibnamefont{Orain}},
  \bibinfo{author}{\bibfnamefont{P.}~\bibnamefont{Maget}},
  \bibinfo{author}{\bibfnamefont{N.}~\bibnamefont{Mellet}},
  \bibinfo{author}{\bibfnamefont{X.}~\bibnamefont{Garbet}},
  \bibinfo{author}{\bibfnamefont{E.}~\bibnamefont{Nardon}},
  \bibinfo{author}{\bibfnamefont{G.}~\bibnamefont{Huysmans}},
  \bibinfo{author}{\bibfnamefont{T.}~\bibnamefont{Casper}},
  \bibinfo{author}{\bibfnamefont{A.}~\bibnamefont{Loarte}},
  \bibinfo{author}{\bibfnamefont{P.}~\bibnamefont{Cahyna}},
  \bibnamefont{et~al.}, \bibinfo{journal}{Nuclear Fusion}
  \textbf{\bibinfo{volume}{52}}, \bibinfo{pages}{054003}
  (\bibinfo{year}{2012}).

\bibitem[{\citenamefont{Furukawa and Zheng}(2009)}]{furukawa09a}
\bibinfo{author}{\bibfnamefont{M.}~\bibnamefont{Furukawa}} \bibnamefont{and}
  \bibinfo{author}{\bibfnamefont{L.-J.} \bibnamefont{Zheng}},
  \bibinfo{journal}{Nuclear Fusion} \textbf{\bibinfo{volume}{49}},
  \bibinfo{pages}{075018} (\bibinfo{year}{2009}).

\bibitem[{\citenamefont{Rutherford}(1973)}]{rutherford73}
\bibinfo{author}{\bibfnamefont{P.~H.} \bibnamefont{Rutherford}},
  \bibinfo{journal}{The Physics of Fluids} \textbf{\bibinfo{volume}{16}},
  \bibinfo{pages}{1903} (\bibinfo{year}{1973}).

\bibitem[{\citenamefont{Waelbroeck et~al.}(2012)\citenamefont{Waelbroeck,
  Joseph, Nardon, B{\'{e}}coulet, and Fitzpatrick}}]{wael12}
\bibinfo{author}{\bibfnamefont{F.}~\bibnamefont{Waelbroeck}},
  \bibinfo{author}{\bibfnamefont{I.}~\bibnamefont{Joseph}},
  \bibinfo{author}{\bibfnamefont{E.}~\bibnamefont{Nardon}},
  \bibinfo{author}{\bibfnamefont{M.}~\bibnamefont{B{\'{e}}coulet}},
  \bibnamefont{and}
  \bibinfo{author}{\bibfnamefont{R.}~\bibnamefont{Fitzpatrick}},
  \bibinfo{journal}{Nuclear Fusion} \textbf{\bibinfo{volume}{52}},
  \bibinfo{pages}{074004} (\bibinfo{year}{2012}).

\end{thebibliography}

\end{document}